\documentclass[a4paper,fleqn,usenatbib]{mnras}

\usepackage{newtxtext,newtxmath}

\usepackage{pbsi}
\usepackage{LobsterTwo}
\usepackage[T1]{fontenc}
\usepackage{ae,aecompl}

\usepackage{graphicx}	
\usepackage{amsmath}	
\usepackage{gensymb}    
\usepackage{hyperref}   
\usepackage{lscape}     
\usepackage[flushleft]{threeparttable} 

\newcommand{\teff}{$T_{\rm eff}$}
\newcommand{\sigteff}{$\sigma(T_{\rm eff})$}
\newcommand{\logg}{$\log g$}
\newcommand{\siglogg}{$\sigma(\log g)$}
\newcommand{\feh}{$\rm{[Fe/H]}$}
\newcommand{\sigfeh}{$\sigma(\rm{[Fe/H]})$}
\newcommand{\vmic}{$v_{t}$}

\newcommand{\qq}{$\mathrm{q}^{2}$}

\newcommand{\sm}{$\rm{M_{\odot}}$}
\newcommand{\sr}{$\rm{R_{\odot}}$}

\newcommand{\be}{\begin{equation}}
\newcommand{\ee}{\end{equation}}
\newcommand{\ben}{\begin{eqnarray}}
\newcommand{\een}{\end{eqnarray}}
\newcommand{\bfg}{\begin{figure}}
\newcommand{\efg}{\end{figure}}

\newcommand{\kepler}{\textit{Kepler}}
\newcommand{\tess}{TESS}
\newcommand{\gaiadrtwo}{\textit{Gaia} DR2}
\newcommand{\gaiaedr}{\textit{Gaia} EDR3}
\newcommand{\prot}{$P_{\rm rot}$}

\newcommand{\inti}{{\LobsterTwo{Inti}}}

\newcommand{\yyisoc}{$Y^{2}$}

\graphicspath{{Plots/}}

\title[]{Searching for new solar twins: The {\LobsterTwo{Inti}} survey for the Northern Sky}

\author[Yana Galarza et al.]
{Jhon Yana Galarza$^{1}$\thanks{E-mail: ramstojh@usp.br},
Ricardo L\'opez-Valdivia$^{2}$,
Diego Lorenzo-Oliveira$^{1}$,
\newauthor
Henrique Reggiani$^{3}$,
Jorge Mel\'endez$^{1}$,
Daniel Gamarra-S\'anchez$^{4}$,
Matias Flores$^{5,6,7}$,
\newauthor
Jerry Portal-Rivera$^{4}$,
Paula Miquelarena$^{5,6,7}$,
Geisa Ponte$^{1,8}$,
Kevin C. Schlaufman$^{3}$,
\newauthor
and Te\'ofilo Vargas Auccalla$^{4}$
\\
$^{1}$Universidade de S\~ao Paulo, Departamento de Astronomia do IAG/USP, Rua do Mat\~ao 1226 \
     Cidade Universit\'aria, 05508-900 S\~ao Paulo, SP, Brazil \\
$^{2}$The University of Texas at Austin, Department of Astronomy, 2515 Speedway, Stop C1400, Austin, TX 78712-1205 \\
$^{3}$ Department of Physics and Astronomy, Johns Hopkins University, 3400 N Charles St., Baltimore, MD 21218 \\
$^{4}$ Seminario Permanente de Astronom\'ia y Ciencias Espaciales, Facultad de Ciencias F\'isicas,
Universidad Nacional Mayor de San Marcos, Avenida Venezuela \\
s/n, Lima 15081, Per\'u \\
$^{5}$ Facultad de Ciencias Exactas, Físicas y Naturales, Universidad Nacional de San Juan, San Juan, Argentina \\
$^{6}$Consejo Nacional de Investigaciones Cient\'ificas y T\'ecnicas (CONICET), Argentina  \\
$^{7}$Instituto de Ciencias Astronómicas, de la Tierra y del Espacio (ICATE), España Sur 1512, CC 49, 5400 San Juan, Argentina \\
$^{8}$CRAAM, Mackenzie Presbyterian University, Rua da Consola\c{c}\~ao, 896, S\~ao Paulo, Brazil
}

\date{Accepted 2021 April 05. Received 2021 March 23; in original form 2020 December 17}

\pubyear{2021}

\begin{document}
\label{firstpage}
\pagerange{\pageref{firstpage}--\pageref{lastpage}}
\maketitle

\begin{abstract}
Solar twins are key in different areas of astrophysics, however only just over a hundred were identified and well-studied in the last two decades. In this work, we take advantage of the very precise \textit{Gaia} (DR2/EDR3), Tycho and 2MASS photometric systems to create the \inti\ survey of new solar twins in the Northern Hemisphere. The spectra of our targets were initially obtained with spectrographs of moderate resolution (ARCES and Goodman spectrographs with $R$ = 31500 and 11930, respectively) to find the best solar twin candidates and then observed at McDonald Observatory with higher resolving power (TS23, $R$ = 60000) and signal-to-noise ratio (SNR $\sim$ 300-500). The stellar parameters were estimated through the differential spectroscopic equilibrium relative to the Sun, which allow us to achieve a high internal precision ($\sigma$(\teff) = 15 K, $\sigma$(\logg) = 0.03 dex, $\sigma$(\feh) = 0.01 dex, and $\sigma$(\vmic) = 0.03 km s$^{-1}$). We propose a new class of stars with evolution similar to the Sun: \textit{solar proxy}, which is useful to perform studies related to the evolution of the Sun, such as its rotational and magnetic evolution. Its definition is based on metallicity ($-$0.15 dex $\leq$ \feh\ $\leq$ +0.15 dex) and mass (0.95 M$_{\odot}$ $\leq$ M $\leq$ 1.05 M$_{\odot}$) constraints, thus assuring that the star follows a similar evolutionary path as the Sun along the main sequence. Based on this new definition, we report 70 newly identified solar proxies, 46 solar analogs and 13 solar-type stars. In addition, we identified 9 \textit{close solar twins} whose stellar parameters are the most similar to those of the Sun.

\end{abstract}



\section{Introduction}
Since the 1980s, the astronomical community was interested in finding stars with physical parameters similar to those of the Sun, i.e., solar twins. However, a crucial question arose: \emph{which parameters should be considered to define an object as solar twin?}. \cite{Strobel:1981A&A....94....1C, Strobel:1989A&A...211..324C, Strobel:1996A&ARv...7..243C} defined these objects as stars whose effective temperature (\teff), surface gravity (\logg), metallicity (\feh), microturbulence (\vmic), photometric properties, chemical composition, age, luminosity, rotation, and magnetic fields are similar, if not identical, to those of the Sun. These authors performed the first attempts to find \textit{real solar twins} exploiting a list of 78 solar analogs obtained by \cite{Hardorp:1978A&A....63..383H} through spectrophotometric observations. However, with these constraints, it was almost impossible to find a \textit{real solar twin}. A less rigorous definition was carried out by \cite{Friel:1993A&A...274..825F}: ``every observable and derivable physical quantity must be identical within observational errors to that of the Sun''. \cite{Porto:1997ApJ...482L..89P}, using the Cayrel de Strobel's definition \citep{Strobel:1989A&A...211..324C}, were able to find the first closest solar twin: 18 Sco. Years later, other authors also claimed to have identified solar twins based on photometric and spectroscopic parameters constraints: HD 143436 \citep{King:2005AJ....130.2318K}, HD 98618 \citep{Melendez:2006ApJ...641L.133M}, HD 10307 and HD 34411 \citep{Galeev:2004ARep...48..492G}, and finally HD 101364 and HD 133600 \citep{Melendez:2007ApJ...669L..89M}, the latter ones not only reproduce the solar fundamental parameters but also a low lithium abundance. On the other hand, \cite{Ramirez:2009A&A...508L..17R} introduced a concept of solar twin based only on spectroscopic stellar parameters constraints (i.e., only \teff, \logg, and \feh), which is useful for achieving high precision differential abundances relative to the Sun; however, this definition introduces a slight bias in the mass-age-\feh\ space that will be discussed later. One of the last definitions comes from \cite{Datson:2012MNRAS.426..484D, Datson:2014MNRAS.439.1028D, Datson:2015A&A...574A.124D}, where a solar twin is defined as a star whose stellar parameters (estimated only with high resolution spectrographs) are indistinguishable from solar within the errors. A very comprehensive discussion about the concept of solar twins and solar analogs is given in \cite{Porto:2014A&A...563A..52P}. These authors suggest that solar twins should not be only indistinguishably from the Sun, but also follow a similar evolutionary history. In brief, the literature is full of different definitions of solar twins, but until now it is not yet clear which parameters should define a \textit{real solar twin}, thus hindering the efforts for finding these objects as well as studies such as gyrochronology and magnetic activity evolution \citep[e.g.,][]{Diego:2018A&A...619A..73L}.

To date, despite different definitions in the literature, approximately 100 solar twins have been identified by different authors \citep[e.g.,][]{Pasquini:2008A&A...489..677P, Melendez:2009ApJ...704L..66M, Melendez:2014A&A...567L...3M, Ramirez:2009A&A...508L..17R, DoNascimento:2013ApJ...771L..31D, Ramirez:2014A&A...572A..48R, Takeda:2009PASJ...61..471T, Baumann:2010A&A...519A..87B, Onehag:2011A&A...528A..85O, Sousa:2011A&A...533A.141S, Gonzales_Hernandez:2010ApJ...720.1592G, Datson:2012MNRAS.426..484D, Porto:2014A&A...563A..52P, Yana_Galarza:2016A&A...589A..65G, Giribaldi:2019A&A...629A..33G} and their applications in different astrophysical fields have had significant impacts on our knowledge about stars and the Sun. For example, they are useful for setting the zero point of fundamental photometric calibrations \citep{Holmberg:2006MNRAS.367..449H, Casagrande:2010A&A...512A..54C, Ramirez:2012ApJ...752....5R, Datson:2014MNRAS.439.1028D, Casagrande:2020arXiv201102517C}, studying the mineralogy of asteroids by subtracting the Sun's reflected light on them \citep[e.g.,][]{Lazzaro:2004Icar..172..179L, Jasmin:2013A&A...552A..85J}, testing stellar interiors through asteroseismology \citep{Chaplin:2011Sci...332..213C, Bazot:2012A&A...544A.106B, Bazot:2018A&A...619A.172B}, measuring distances \citep{Jofre:2015MNRAS.453.1428J}, and even improving spectroscopic methods for stellar parameters determination \citep{Saffe:2018A&A...620A..54S}. More recently, \citet{Yana:2021MNRAS.502L.104Y} detected for the first time a differential odd-even effect in a solar twin, providing new insights to understand the supernova nucleosynthesis history.

The study of solar twins has also contributed to the understanding of the chemical evolution of the Galactic disk \citep{daSilva:2012A&A...542A..84D, Nissen:2015A&A...579A..52N, Spina:2016A&A...593A.125S, Spina:2018MNRAS.474.2580S, Bedell:2018ApJ...865...68B, Nissen:2020A&A...640A..81N, Bothelho:2020MNRAS.499.2196B}, and the study of the neutron-capture elements \citep{Melendez:2014ApJ...791...14M, Yana_Galarza:2016A&A...589A..17Y}. As a result, new chemical clocks as the [Y/Mg]-age correlation initially proposed by \citet{daSilva:2012A&A...542A..84D} and improved by several authors \citep[e.g.,][]{Nissen:2015A&A...579A..52N, Tucci:2016A&A...590A..32T, Feltzing:2017MNRAS.465L.109F, Spina:2018MNRAS.474.2580S, Jofre:2020A&A...633L...9J, Nissen:2020A&A...640A..81N} and the Li-age correlation \citep{Baumann:2010A&A...519A..87B, Carlos:2016A&A...587A.100C, Carlos:2019MNRAS.tmp..667C} have been established. Biology principles (Phylogenetics of solar twins) have also been applied to investigate the chemical evolution of the Milky Way \citep{Jofre:2017MNRAS.467.1140J, Holly:2020arXiv201106453J}.

Significant contributions come from the works of \citet{Nascimento:2013ApJ...771L..31D, DoNascimento:2020ApJ...898..173D} and \cite{Diego:2019MNRAS.485L..68L, Diego:2020MNRAS.495L..61L} in the field of gyrochronology using solar twins, giving important clues to understand the rotational evolution of the Sun. A controversial diagnostic of stellar ages is the age-chromospheric activity relation \citep{Mamajek:2008ApJ...687.1264M, Pace:2013A&A...551L...8P}, whose applicability is extended for stars with $\sim$6-7 Gyr \citep{Diego:2016A&A...594L...3L}. Besides, the analysis of solar twins could help us to place the $\sim$11 yr solar cycle in context \citep{Hall:2007AJ....133..862H, Hall:2009AJ....138..312H, Flores:2018A&A...620A..34F}. \cite{Melendez:2009ApJ...704L..66M} reported chemical anomalies in the Sun when it is compared to solar twins, thereby establishing the basis for studying the planet-stellar chemical composition connection \citep[e.g.,][]{Ramirez:2009A&A...508L..17R, Gonzales_Hernandez:2010ApJ...720.1592G, Gonzales_Hernandez:2013A&A...552A...6G, Schuler:2011ApJ...732...55S, Adibekyan:2014A&A...564L..15A, Maldonado:2015A&A...579A..20M, Spina:2016A&A...593A.125S, Nissen:2015A&A...579A..52N, Nissen:2017A&A...608A.112N, Fan_Liu:2016MNRAS.456.2636L, Bedell:2018ApJ...865...68B, Tucci:2019A&A...628A.126M, Cowley:2020RNAAS...4..106C}. Finally, the study of solar twins is also expanded to the field of exoplanets. For instance, \citet{Bedell:2015A&A...581A..34B, Bedell:2017ApJ...839...94B} and \cite{ Melendez:2017A&A...597A..34M} have demonstrated that with the high precision achieved in stellar parameters in solar twins, it is possible to get very precise exoplanet properties (mass and radius). Additionally, it is also feasible to study habitability and evolution of exoplanets through radioactive elements such as thorium \citep[e.g.,][]{Unterborn:2015ApJ...806..139U, Botelho:2019MNRAS.482.1690B}.

As widely discussed above, the identification of new solar twins is essential to the advancement of diverse astronomical fields. Therefore, in this work we present the \inti\footnote{\inti\ means Sun in the Inca-Andean-Quechua cosmovision.} catalog of new solar twins, solar analogs, and solar-type stars identified in the Northern Hemisphere. The \inti\ survey provides reliable and precise spectroscopic stellar parameters, chromospheric activity levels, and photometric rotational periods when its determinations through high-precision light curves are possible.

\begin{table}
	\centering
	\caption{\textit{Gaia} Absolute Magnitude ($M_{G}$) and Photometric color constraints established using known solar twins \citep{Ramirez:2014A&A...572A..48R}. $V_{TG}$ is the $V$ magnitude based on Eq. (\ref{eq:1}), while $B_{T}$ represent the Tycho $B$ magnitude, $G_{\rm{BP}}$ and $G_{\rm{RP}}$ are \gaiaedr\ magnitudes in the BP and RP passbands, and $J$, $H$, $K_{S}$ are 2MASS magnitudes.}
	\label{tab:1}
	\begin{tabular}{c} 
		\hline
		\hline
		\textit{Gaia} Absolute Magnitude \& color constraints                                   \\
		\hline
		$3.755  \leq M_{G}                   \leq 5.331$     \\
		$0.254  \leq G_{\rm{BP}}-G           \leq 0.377$     \\
		$0.455  \leq G-G_{\rm{RP}}           \leq 0.589$     \\
		$0.761  \leq G_{\rm{BP}}-G_{\rm{RP}} \leq 0.907$     \\
		$0.960  \leq G-J                     \leq 1.315$     \\
		$1.207  \leq G-H                     \leq 1.708$     \\
        $1.284  \leq G-K_{S}                 \leq 1.791$     \\
        $0.166  \leq J-H                     \leq 0.506$     \\
        $0.281  \leq J-K_{S}                 \leq 0.550$     \\
        $-0.002 \leq H-K_{S}                 \leq 0.180$     \\
        $0.473  \leq B_{T}-V_{TG}             \leq 0.947$     \\
		\hline
		\hline
	\end{tabular}
\end{table}


\begin{figure}
 \includegraphics[scale=0.75]{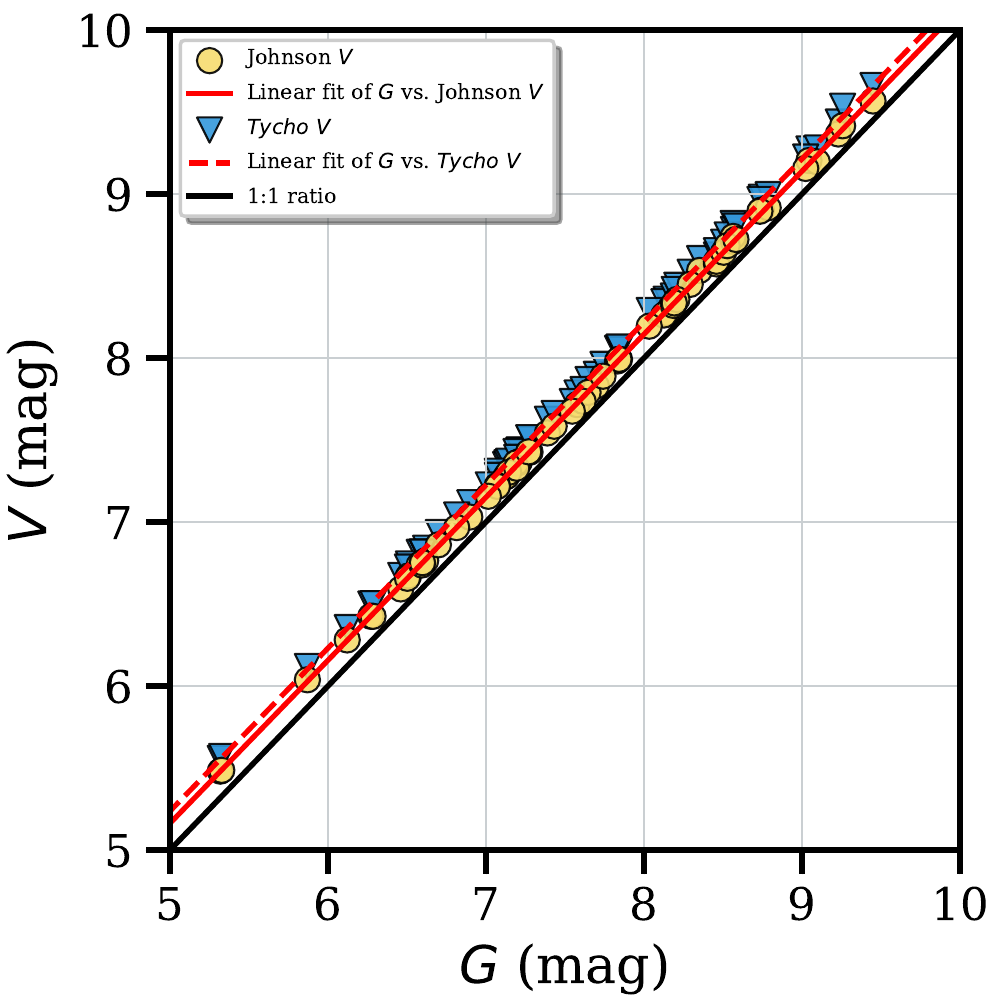}
 \centering
 \caption{Photometric relationship between Johnson/Tycho $V$ vs. \gaiaedr\ $G$ magnitudes, which are valid only for solar twins. Both the red line and the red dashed lines represent the linear fit to the data considering errors in both axis, while the black line is the 1:1 relation. Error bars are smaller than the symbols.}
 \label{fig:V_to_G_final}
\end{figure} 

\section{Sample Selection}
Inspired by seminal works for searching solar twins \citep[e.g.,][]{Hardorp:1978A&A....63..383H, Strobel:1981A&A....94....1C, Melendez:2007ApJ...669L..89M, Ramirez:2009A&A...508L..17R, Porto:2014A&A...563A..52P}, in this era of large surveys, we take advantage of the precise photometric systems of \textit{Gaia} DR2/EDR3 \citep{GAIADR2:2018A&A...616A...1G, GAIADR3:2020arXiv201201533G}, Tycho \citep{Hipparcos_tycho:2000A&A...355L..27H}, and 2MASS \citep{2mass:2006AJ....131.1163S} to search these stellar objects. However, unlike these initial studies mainly based on photometric comparisons with the Sun, our methodology consist in performing color constraints from the well characterized solar twins of \cite{Ramirez:2014A&A...572A..48R}. To do so, we cross-matched the solar twin sample with the TGAS \citep[Tycho-Gaia Astrometric Solution,][]{GAIADR1:2017A&A...607A.105M} catalog updated with the \gaiaedr\ magnitudes, resulting in 63 common objects. In our study we also found a simple photometric relationship from \gaiaedr\ $G$ (updated from our initial relations using DR2) to Tycho $V$ and Johnson $V$ \citep{Kharchenko:2001KFNT...17..409K}:
\ben
\label{eq:1}
V_{TG} &=& G \times 0.9942 (\pm0.0021) + 0.2657 (\pm0.0156) \\
V_{G}  &=& G \times 0.9940 (\pm0.0014) + 0.1929 (\pm0.0097),
\label{eq:2}
\een
where $V_{TG}$ is the transformation between $G$ and Tycho V, while $V_{G}$ is the conversion between $G$ and Johnson $V$ (see Figure \ref{fig:V_to_G_final}). The dispersion and the reduced chi-squared ($\chi^{2}_{red}$) of the linear fit are 0.018 mag and 1.608 for Eq. (\ref{eq:1}), and 0.014 mag and 1.554 for Eq. (\ref{eq:2}), respectively. It is important to highlight that these relationships are valid only for solar twins\footnote{A similar relationship was found between \gaiadrtwo\ $G$ and Johnson $V$: $V_{G} = G \times 0.9901 (\pm0.0014) + 0.2346 (\pm0.0097)$}. Besides, $V_{TG}$ is used only to establish the color constraints showed in Table \ref{tab:1}, while $V_{G}$ is useful to estimate isochronal ages as it will be discussed later. In this way, we established the bounds of our color constraints, which are shown in Table \ref{tab:1}.

\begin{figure}
 \includegraphics[width=\columnwidth]{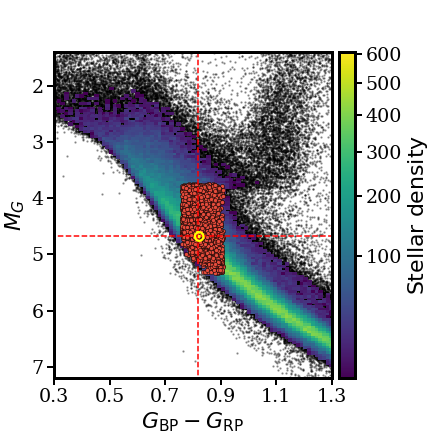}
 \centering
 \caption{\gaiaedr\ HR diagram of $\sim$1.4 million stars within 200 pc from the Solar System (the data were obtained using the \texttt{astroquery} package). The colormap represents the stellar density distribution created from the 2D-histogram function of the matplotlib library. The red circles are the solar twin candidates found after applying our color constraints. The Sun is plotted as a yellow solar standard symbol.}
 \label{fig:MG_BPRP}
\end{figure}

The sample selection technique consists in first cross-matching the \gaiadrtwo\ (updated to EDR3 in Table \ref{tab:1}) with the 2MASS and Tycho catalogs within a region of 100 pc from Earth and with $G$ values ranging from 5 to 9 mag. Then, we applied our color constraints in the cross-matched sample and found 3100 objects. These objects compose our preliminary sample of solar twin candidates and are plotted as red circles in the \gaiaedr\ Hertzsprung-Russell diagram\footnote{Our HR diagram is based on \url{https://vlas.dev/post/gaia-dr2-hrd/}} of Figure \ref{fig:MG_BPRP}. We did not apply reddening corrections for our sample, since it is within 100 pc and thus has negligible reddening. This is supported by \citet{Vergely:2010A&A...518A..31V} and \citet{Lallement:2014A&A...561A..91L}, who found a gradient of d E(B-V)/dr = 0.0002 mag per pc (see Figure 2 in \citealp{Lallement:2014A&A...561A..91L}). A similar result is found by \citet{Green:2019ApJ...887...93G}, who created a precise dust map using the Pan-STARRS1, 2MASS, and \gaiadrtwo\ (including its parallaxes) photometric bands. On the other hand, \citet{Reis:2011ApJ...734....8R} reported interstellar absorption (E(B-V) $>$ 0.056) in the local bubble for regions on the Galactic plane ($d>60$ pc) with latitudes from $l\geq270^{\circ}$ up to $l\leq45^{\circ}$. However, the stars of the \inti\ survey that fall in this region have $d<50$ pc and thus E(B-V) = 0.

Finally, as our photometric methodology (see Table \ref{tab:1}) for searching solar twins is based only on stellar parameter constraints \citep{Ramirez:2009A&A...508L..17R}, it is expected to have a bias in our results (see the irregular polygon in Figure \ref{fig:MG_BPRP}), as our criteria are not symmetric relative to the main sequence evolution of a one-solar-mass solar-metallicity star. This will be discussed more extensively in the subsection \ref{subsec:new solar twins}.

\section{Spectroscopic observations and data reduction}
\label{sub:observations and data reduction}
We obtained spectra of high signal-to-noise ratio (SNR $\sim400$) of our candidates with the Goodman High Throughput Spectrograph \citep{Clemens:2004SPIE.5492..331C}, on the SOAR Telescope. From these spectra, we could retrieve a reliable initial guess of the spectroscopic parameters. The methodology consists in estimating \feh\ through spectral synthesis with the SP\_ACE spectral analysis tool\footnote{\url{http://dc.g-vo.org/SP_ACE}}, \teff\ by using the colour–temperature–metallicity calibrations established by \citet{Casagrande:2010A&A...512A..54C}, and trigonometric \logg\ from \gaiaedr\ parallaxes with bolometric corrections given in \citet{Melendez:2006ApJ...641L.133M}. Our technique was tested using several known solar twins and the results are consistent (within the uncertainties) with those estimated with high precision spectra \citep[e.g.,][]{Spina:2018MNRAS.474.2580S}. We also used the ARC Echelle Spectrograph (ARCES) on the 3.5-meter Apache Point Observatory telescope to explore the northern sky since the observations with the SOAR telescope are inaccessible to DEC $\gtrsim$ +25$^{\circ}$. Thanks to these initial observations, we created a sample of 150 objects with the best solar twin candidates which were later observed with the Robert G. Tull Coud\'e Spectrograph \citep[hereafter TS23,][]{Tull:1995PASP..107..251T} on the McDonald Observatory. The number of stars observed with each instrument and the internal precision achieved in stellar parameters are summarized in Table \ref{tab:2}. In the following are detailed the spectroscopic observations performed in each observatory.

\subsection{SOAR Telescope at Cerro Pachon}
The spectra of the first potential solar twin candidates were obtained using the Goodman spectrograph on the 4.1-meter SOAR Telescope under the programs SO2017B-004, SO2018A-005, SO2019A-007, and SO2019B-005 from 2017 to 2019. The solar spectrum was obtained after a short exposure of the Moon. The instrument was configured to use the red camera and the grating of 2100 lines/mm, resulting in a moderate resolving power $R = \lambda / \Delta \lambda = 12000$ and wavelength coverage of 630 \AA\ centered at H$\alpha$. The Goodman spectra were reduced using IRAF\footnote{IRAF is distributed by the National Optical Astronomy Observatory, which is operated by the Association of the Universities for Research in Astronomy, Inc. (AURA) under cooperative agreement with the National Science Foundation.} following the standard procedure, i.e., creation of the master flat, flat field correction, sky subtraction, order extraction, etc. The radial velocity correction was performed using the \texttt{rvidlines} and \texttt{dopcor} task of IRAF. The obtained spectra were also normalized using IRAF's \texttt{continuum} task with orders ranging from two to five.

\subsection{Apache Point Observatory}
We also obtained the spectra of the solar twin candidates using the ARCES on the 3.5-meter telescope at the Apache Point Observatory. The observations were carried out from 2019 to 2020. The solar spectra were obtained by observing the sky at twilight time. We used the CERES\footnote{\url{https://github.com/rabrahm/ceres}} pipeline in order to perform the standard reduction of the ARCES spectra. The ARCES spectrograph provides spectra of $R \sim 31500$ and covers the entire visible wavelength (from 3200-10000 \AA). The SNR achieved ranges from 200 to 300 at $\sim$6000 \AA.

\subsection{McDonald Observatory} 
All the observations were taken during the years of 2018-2020 using the TS23 configured in its high resolution mode on the 2.7-meter Harlan J. Smith Telescope at McDonald Observatory. The spectra of the Sun were obtained through the reflected light from the Moon. As the McDonald Observatory does not have an official pipeline to reduce the TS23 spectra, we have developed our own scripts\footnote{\label{github}\url{https://github.com/ramstojh}} based on the practical reduction notes of the Dr. Chris Sneeden, Dr. Ivan Ram\'irez, and Dr. Diego Lorenzo-Oliveira. The code consists of a number of semi-automatic python scripts that performs bias subtraction, flat fielding, order extraction, and wavelength calibration using PyRAF \citep{PyRAF:2012ascl.soft07011S}. The resulting spectra are free of fringing defects and have $R = 60000$, $\rm{SNR} \sim $300-500 at $\sim$6500 \AA, and cover a wide spectral range (from 3750-9900 \AA).

\begin{table}
	\centering
	\caption{Number of stars observed with each instrument and the internal precision achieved in stellar parameters. ($\star$) Number of stars observed in the Northern Hemisphere with -18$^{\circ}$ $\leq$ DEC $\leq$ +25$^{\circ}$. }
	\label{tab:2}
	\begin{tabular}{ccccc} 
		\hline
		\hline
		Instrument & Observed stars & \sigteff & \siglogg & \sigfeh    \\
		           &                &   (K)    &   (dex)  &  (dex)     \\
		\hline
		Goodman    &  160$^{\star}$  & 190      & 0.20     & 0.15       \\
		ARCES      &  20            & 100      & 0.11     & 0.08       \\
		TS23       &  147           & 15       & 0.03     & 0.01       \\
		\hline
		\hline
	\end{tabular}
\end{table}


\begin{figure*}
 \includegraphics[scale=0.35]{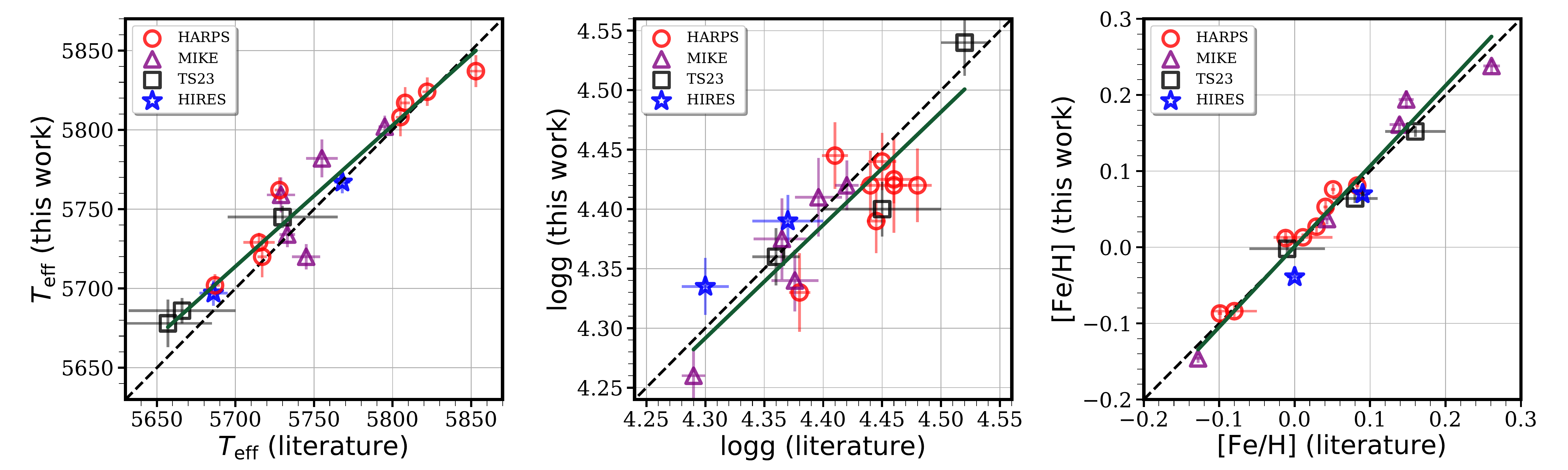}
 \centering
 \caption{Comparison between our stellar parameters and those obtained by \citet{Ramirez:2013ApJ...764...78R, Ramirez:2014A&A...572A..48R} (squares and triangles) and \citet{Spina:2016A&A...585A.152S, Spina:2018MNRAS.474.2580S} (stars and circles). The dashed lines represent the 1:1 ratio, while the green lines are the linear fits for \teff\ (left panel, $rms = 12$ K), \logg\ (middle panel, $rms = 0.03$ dex), and \feh\ (right panel, $rms = 0.02$ dex).}
 \label{fig:sp_comparison_old_st}
\end{figure*}

\section{Fundamental parameters}
\label{sub:fundamental parameters}
\subsection{Equivalent Widths and Stellar Parameters}
In order to perform the standard treatment and analysis of the TS23 spectra, we have developed semi-automatic python scripts\textsuperscript{\ref{github}}, whose structure is a combination of two tools: iSpec\footnote{\url{https://www.blancocuaresma.com/s/iSpec}} and IRAF. In summary, the scripts use the iSpec tool to perform the radial/barycentric velocity correction, and the \texttt{continuum} and \texttt{scombine} tasks of IRAF to normalize and combine the TS23 spectra. All the processes mentioned above are automatic resulting in spectra of high quality and SNR ($\sim$300-500 at 6500 \AA). The scripts are also capable of measuring Equivalent Widths ($EW$s) through Gaussian fits to the line profile using the Kapteyn \texttt{kmpfit} Package\footnote{\url{https://www.astro.rug.nl/software/kapteyn/}} in windows of 6 \AA; however, this process is manually performed in order to achieve a higher precision. The method is based on line-by-line equivalent width measurements between the Sun and the object of interest, choosing consistent pseudo-continuum regions for both objects \citep[e.g.,][]{Melendez:2009ApJ...704L..66M, Bedell:2014ApJ...795...23B, Yana_Galarza:2016A&A...589A..17Y, Spina:2018MNRAS.474.2580S}. Besides, the script generates an output file containing information about the local continuum, limits of the Gaussian fits, $\chi^{2}$ test, excitation potential, oscillator strength, and laboratory $\log (gf)$ values \citep[see][]{Melendez:2014ApJ...791...14M}. On the other hand, as the ARCES spectra are already corrected by radial velocity shifts, we used our python scripts only to measure the $EW$s, rigorously following the same procedure already explained above. 

As in our previous works \citep[e.g.,][]{Ramirez:2014A&A...572A..48R, Yana:2019MNRAS.490L..86Y}, we employed the automatic \qq\ (qoyllur-quipu)\footnote{\url{https://github.com/astroChasqui/q2}} python code to determine the spectroscopic stellar parameters (\teff, \logg, \feh, \vmic) for our sample. In short, the code estimates the iron abundances using the line list from \cite{Melendez:2014ApJ...791...14M} and the 2019 version of the local thermodynamic equilibrium (LTE) code MOOG \citep{Sneden:1973PhDT.......180S} with the Kurucz ODFNEW model atmospheres \citep{Castelli:2003IAUS..210P.A20C}. Then, the \qq\ employs the spectroscopic equilibrium, which is a standard technique of iron line excitation and ionization equilibrium. As a result, we obtain very reliable stellar parameters with high internal precision $\sigma$(\teff) = 15 K, $\sigma$(\logg) = 0.03 dex, $\sigma$(\feh) = 0.01 dex, and $\sigma$(\vmic) = 0.03 kms$^{-1}$. The masses were inferred from an isochrone analysis, which is described in detail in subsection \ref{subsec:age and mass}. Our inferred stellar parameters can be found in Table \ref{appendix_tab:1}. In order to test the precision of the scripts and the reliability of the results, we compared our stellar parameters with those from \cite{Ramirez:2013ApJ...764...78R, Ramirez:2014A&A...572A..48R} and \cite{Spina:2016A&A...585A.152S, Spina:2018MNRAS.474.2580S}. As shown in Figure \ref{fig:sp_comparison_old_st}, there is a good agreement between our results and those obtained using spectrographs of even higher resolution (e.g., HARPS spectrograph with $R \sim 115000$ in \citet{Spina:2018MNRAS.474.2580S}) than the TS23. 

\subsection{New solar twins}
\label{subsec:new solar twins}
As discussed earlier, the concept of solar twins changed over time \citep[e.g.,][]{Strobel:1981A&A....94....1C, Friel:1993A&A...274..825F, Datson:2012MNRAS.426..484D, Ramirez:2014A&A...572A..48R}, and most of them are generally based on photometric and spectroscopic stellar parameters constraints but not on fundamental parameters that drive evolutionary states. The lack of the latter introduces a bias in the mass of the known solar twins, i.e., most of them are slightly more massive ($\sim$ 0.03 \sm) than the Sun \citep[][]{Ramirez:2014A&A...572A..48R, Spina:2018MNRAS.474.2580S}. Thus, hampering the sample selection for studies such as gyrochronology and magnetic activity evolution \citep[e.g., ][]{Diego:2018A&A...619A..73L, Diego:2019MNRAS.485L..68L}. It is expected that our sample also has the same bias because we used the stellar parameters constraints given by \cite{Ramirez:2014A&A...572A..48R} (i.e., all stars with stellar parameters into \teff\ = 5777 $\pm$ 100 K, \logg\ = 4.44 $\pm$ 0.10 dex, \feh\ = 0.00 $\pm$ 0.10 dex) in our solar twin hunting program. Figure \ref{fig:MG_BPRP} clearly shows that these constrains (in shape of an irregular polygon) do not follow the evolutionary tracks of main sequence stars, thereby introducing a slight bias in mass and removing several solar twin candidates with M $<$ 1.0 M$_{\odot}$ in our sample.

\begin{figure*}
 \includegraphics[scale=0.4]{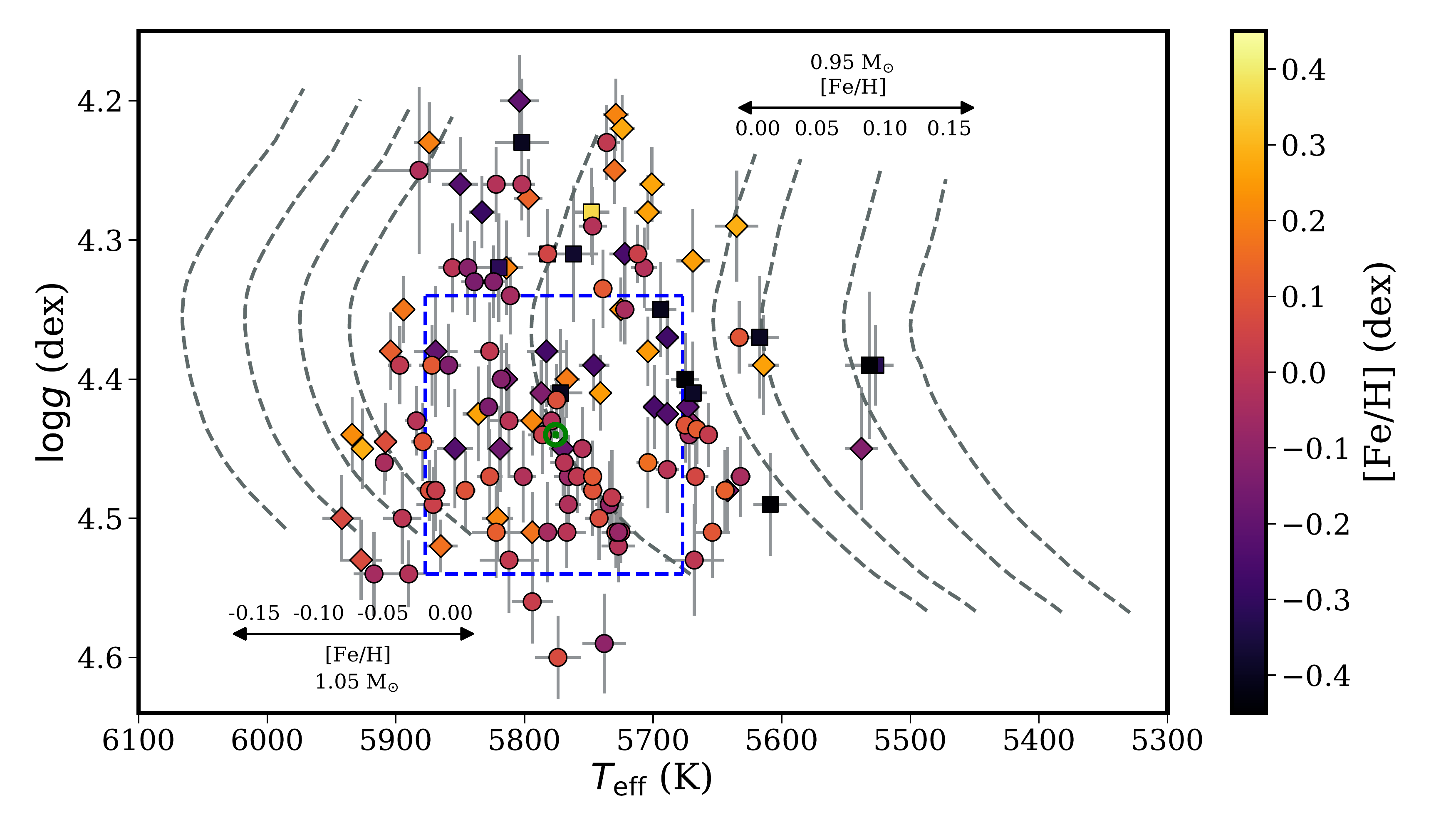}
 \centering
 \caption{Hertzsprung-Russell diagram plotted using the evolutionary tracks of $Y^{2}$ \citep{Yi:2001ApJS..136..417Y, Demarque:2004ApJS..155..667D}. The left dashed lines represent the evolutionary tracks for 1.05 M$_{\odot}$ with \feh\ from 0.00 to -0.15 dex, while the right dashed lines for 0.95 M$_{\odot}$ with \feh\ from 0.00 to 0.15 dex. The new solar twins and solar proxies are plotted as circles. To differentiate solar twins from solar proxies we also plotted the Ramirez's solar twin definition in blue dashed lines, i.e., all the circles plus their error bars that fall within this region are considered as solar twins. The} solar analogs and the solar-type stars are shown in diamonds and squares, respectively. The colormap represents the \feh\ for all the stars. The Sun's data is plotted as reference (green solar standard symbol) with its evolutionary track (i.e., dashed line for 1.0 M$_{\odot}$ and \feh\ = 0.0 dex).
 \label{fig:McDonald_nst}
\end{figure*}

In order to address this issue for studies of the Sun's evolution along the main sequence, such as rotational and magnetic evolutionary studies, we propose a new class of star like the Sun\footnote{Historically there are three classes of stars like the Sun: solar twins, solar analog stars, and solar-type stars \citep{Strobel:1996A&ARv...7..243C}.}: \textit{solar proxy stars}. Its definition is based on spectroscopic stellar parameter constraints ensuring that the star follows an evolutionary path similar to the Sun on the main sequence. Such constraints are: 1) \feh\ and mass values within $\pm$0.15 dex and $\pm$5\% of the Sun's, respectively. Stars with these values roughly follow a similar evolutionary path as the Sun; 2) the \logg\ is only constrained to verify if the star is on the main sequence and its bounds depend on the mass isochrone model (e.g., \logg\ can takes values from $\sim$4.1 dex to 4.6 dex for a solar-mass star); 3) as a result of the above constraints, the \teff\ takes larger values ranging from $\sim$5320 K to $\sim$6050 K, nonetheless to achieve a high precision in stellar parameters we recommend to use \teff\ values within 150-200 K of the solar value (e.g., to avoid differential 3D and non-LTE effects). All these constraints are very well represented in Figure \ref{fig:McDonald_nst}, where the dashed lines are the evolutionary tracks taken from Yonsei-Yale ($Y^{2}$) isochrones \citep{Yi:2001ApJS..136..417Y, Demarque:2004ApJS..155..667D} for masses between 0.95 M$_{\odot}$ and 1.05 M$_{\odot}$, in steps of 0.05 dex in \feh. Solar analog and solar-type stars keep the same definition, i.e., solar analogs are objects with \feh\ within a factor of two of solar \citep{Soderblom:1998saco.conf...41S}, while solar-type stars are main sequence or subgiant stars with 5000 K $<$ \teff\ $<$ 6500 K or spectral type ranging from F8V to K2V as is defined in the literature \citep[e.g., see][]{Soderblom:1998saco.conf...41S, Adibekyan:2017AN....338..442A}.

Applying the above definitions in our sample, we identified 70 solar proxies (from which 42 are solar twins according to the definition of \citealp{Ramirez:2014A&A...572A..48R}), 46 solar analogs, and 13 solar-type stars, which are represented by circles (solar twins and proxies), diamonds, and squares, respectively in Figure \ref{fig:McDonald_nst}. It is important to mention that we consider as solar twin/proxy/analog star to those whose uncertainties in fundamental parameters fall into our definition criteria, since our precision is limited by the resolving power. In addition, there are 23 wide binaries \citep{Tokovinin:2014AJ....147...86T, Tokovinin:2014AJ....147...87T} from which 14 are solar proxies, 7 are solar analogs and 2 are solar-type stars. The spectra of these binaries are not contaminated by their companions. We also identified four new spectroscopic binaries which are not analyzed in this work, however they are summarized in Table \ref{appendix_tab:4}. Despite some stars of our sample have been already analyzed by other authors, their stellar parameters were estimated using methods that are different from ours, and with lower SNR or resolving power, or different spectral coverage. In order to identify the \textit{closest solar twin} in our sample, we have narrowed down the \feh, \teff, \logg, and mass to be within of 0.05 dex, 50 K, 0.05 dex, and 0.03 \sm\ of the solar values. As a result, 9 stars (HIP 49580, HIP 20218, HIP 11253, HIP 7244, HD 105590A, HD 49425, HD 22875, HD 9201, TYC 1678-109-1) met these strict criteria, whose solar masses and \feh\ make them useful for obtaining precise chemical abundances differentially to the Sun. These few objects represent 7\% from the total of stars analyzed in this work, thus showing how hard they are to find.

\subsection{Age and Mass}
\label{subsec:age and mass}
Isochronal mass and age determinations helped us to better understand the evolution of stars, calibrate age correlations as gyrochronology and magnetic activity evolution, as well as to understand the Chemical Evolution of the Galaxy. However, this method relies on isochrones of stellar evolution models and input stellar parameters (\teff, absolute magnitude ($M_{V}$) and \feh) \citep[e.g.,][]{Lachaume:1999A&A...348..897L, Takeda:2007A&A...468..663T} that usually give large uncertainties ($\sim$3-4 Gyr), biases, or sometimes serious spurious age results because the $M_{V}$ is generally estimated from photometry of moderate precision. In order to increase the precision in the method, \cite{Ramirez:2013ApJ...764...78R, Ramirez:2014A&A...572A..48R} replaced the $M_{V}$ by the spectroscopic \logg. Using the \qq\ code, that performs probability distribution functions of ages and masses from the \yyisoc\ isochrones, these authors not only were able to achieve precise age values, but also greatly reduce their uncertainties to $\sim$1-2 Gyr \citep[e.g., see the compelling example of HIP 56948 in][]{Melendez:2012A&A...543A..29M}. \cite{Spina:2018MNRAS.474.2580S} improved the \qq\ interpolation method by including $\alpha$-enhancements, spectroscopic \logg, and $M_{V}$  (estimated through Hipparcos/\textit{Gaia} parallaxes) as input parameters (see their Eq. 3). As a result, these authors further reduce the uncertainties to $\sim$1.0 Gyr ($\sim$0.5 Gyr using \textit{Gaia} DR2 parallaxes). Despite the improvements performed by \citet{Ramirez:2014A&A...572A..48R} and \cite{Spina:2018MNRAS.474.2580S} to the isochronal age estimator, there is still a strong dependency on the precision of the spectroscopic \logg\, which at the same time is sensitive to $EW$s measurements and therefore also to the ionization balance. 

\begin{figure*}
 \includegraphics[scale=0.38]{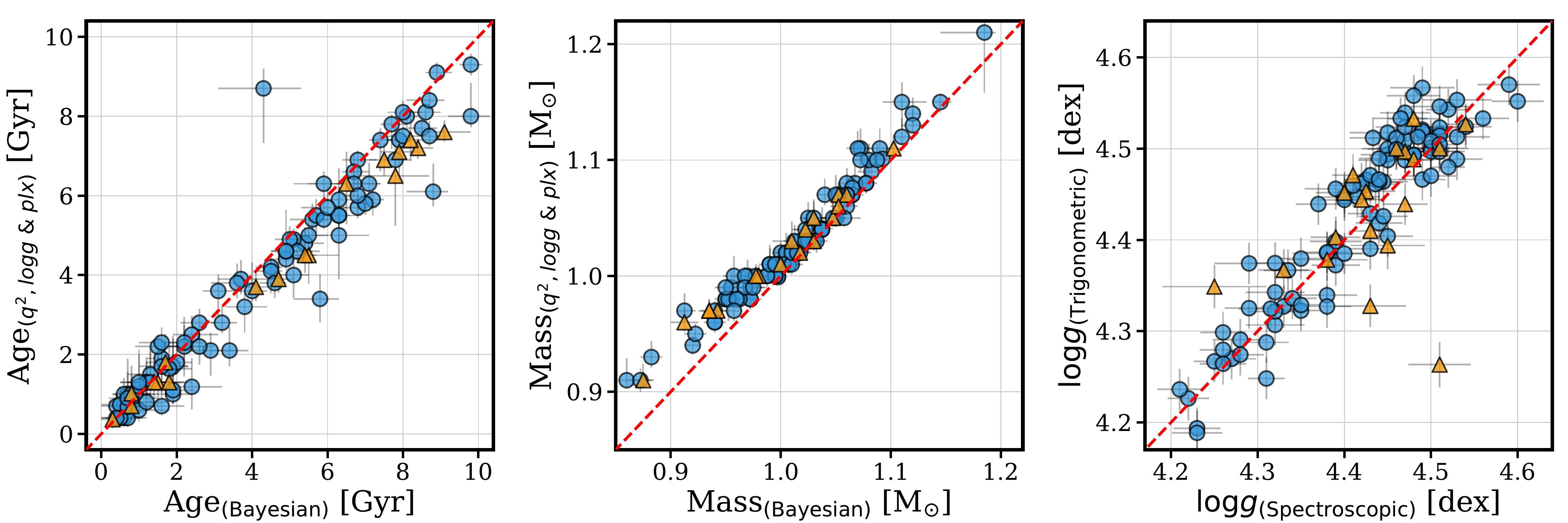}
 \centering
 \caption{\textbf{Left panel:} Comparison between the \qq\ ages estimated using \logg\ \& parallaxes as input parameters versus the Bayesian ages. The latter includes \textit{Gaia} DR2 $G$ magnitudes in its calculations. \textbf{Middle panel:} \qq\ (\logg \ \& parallaxes) masses versus Bayesian masses. \textbf{Right panel:} \textbf{Comparison between} the spectroscopic and the trigonometric \logg\ (Eq. (\ref{eq:3})). In all panels, the dashed red lines represent the 1:1 ratio, while the yellow triangles the binary stars.}
 \label{fig:fp_comparison_st_ages_logg}
\end{figure*} 

In this work, we estimate isochronal ages using both the improved \qq\ and the Bayesian inference. The improved \qq\ uses stellar parameters, parallaxes and Johnson $V$ magnitudes as input parameters. In our age estimations we use the precise \gaiaedr\ parallaxes, which are corrected by subtracting $-$15 $\pm$ 18 $\mu$as as suggested by \citet{Stassun:2021arXiv210103425S}. Besides, as the Johnson $V$ is not as precise as Gaia $G$, we used the $V_{G}$ (estimated from Eq. (\ref{eq:2}) and with $\sigma(V_{G})\sim0.015$) when the uncertainties of Johnson $V$ are greater than 0.015. It helps us to improve the q2 age estimations and it is also useful when Johnson $V$ is not available. We found a dispersion of only $\sim$0.04 Gyr between the \qq\ ages estimated using Johnson $V$ and $V_{G}$. On the other hand, the Bayesian inference method \citep{Grieves:2018MNRAS.481.3244G} employs the \yyisoc evolutionary tracks adopting steps of 0.01 M/$\rm{M_{\odot}}$ in mass, 0.05 dex in metallicity, and 0.05 dex in [$\alpha$/Fe]. Posterior distributions of ages and other evolutionary parameters are estimated through the proper marginalization of the likelihood as a function of \teff, \feh\footnote{In this work, we adopted $[\alpha$/Fe] = 0 as we did not determine the abundance of $\alpha$-enhancement elements.}, \logg, \gaiaedr\ parallaxes (already offset by $-$15 $\mu$as) and \gaiadrtwo\ $G$ band photometry. We emphasize that we did not use \gaiaedr\ G band in the Bayesian method as there are not yet bolometric corrections ($BC$) for it. For the brightest stars ($G < 6$ mag), we corrected the \gaiadrtwo\ $G$ band systematics and applied the $BC$ of \citet{Casagrande:2018MNRAS.479L.102C} to estimate luminosities. The resulting photometric errors are composed by the quadratic propagation of the nominal $G$ band errors reported by \gaiadrtwo\ and a conservative lower limit of 0.01 mag. The likelihood function is evaluated along each possible evolutionary step (within $\pm$10$\sigma$ of the input parameter space) and simultaneously weighted by metallicity and mass inputs, which are based on the solar neighborhood metallicity distribution \citep{Casagrande:2018IAUS..330..206C} and Salpeter initial mass function, respectively. The values adopted for each one of the evolutionary parameters result from the median (50\% percentile) and $\pm1\sigma$ intervals (16-84\% percentile) yielded by its posterior cumulative distributions.

The left panel of Figure \ref{fig:fp_comparison_st_ages_logg} shows the \qq\ ages estimated for our sample using spectroscopic stellar parameters and parallaxes as input parameters (hereafter \logg\ \& plx) versus the Bayesian inference ages. There is a good agreement between methods with a dispersion of only $\sim$0.48 Gyr. This dispersion is estimated removing the most prominent outlier, which is a star with \feh = $-$0.5 dex and M = 0.85 \sm. In the middle panel of Figure \ref{fig:fp_comparison_st_ages_logg}, we compare the masses and it is shows good agreement with a dispersion of only 0.01 \sm. As the spectroscopic \logg\ is a fundamental observable for estimating isochronal ages, it is important to make a comparison with the trigonometric gravity. It is estimated from the luminosity ($\propto R^{2}$\teff$^{4}$) and gravity ($\propto$ M/R$^{2}$) relations to arrive to the following expression:
\be
\begin{split}
\label{eq:3}
\log g (Trig) = \log \left(\frac{\rm{M}}{\rm{M}_{\odot}}\right) + 4 \log \left(\frac{T_{\rm eff}}{T_{\odot}}\right) + 0.4 V + 0.4 BC & \\
+ 2 \log \left(\frac{plx}{1000}\right) + 0.104,
\end{split}
\ee
where $V$ is the unreddened visual magnitude, $plx$ the parallax in milliarcsecond, and $BC$ the bolometric correction. The latter is taken from \citet{Melendez:2006ApJ...641L.133M}. The last term of Eq. (\ref{eq:3}) is somewhat different from the literature \citep[e.g.,][]{Nissen:1997ESASP.402..225N} because we adopted a slightly different absolute bolometric magnitude value for the Sun ($M_{bol,\odot}$ = 4.74; \citealp{Bessell:1998A&A...333..231B}). As can be seen in the right panel of Figure \ref{fig:fp_comparison_st_ages_logg}, there is generally good agreement between the spectroscopic and trigonometric \logg (almost within $2\sigma$) with a dispersion of 0.035 dex. Binary stars (represented by yellow triangles) are not considered into the dispersion estimation in all panels of Figure \ref{fig:fp_comparison_st_ages_logg}. Therefore, with the above results, we can conclude that the ages determined using \qq\ are as precise as those from the Bayesian inference. This is somewhat expected since the errors derived in this work are small enough to reduce the importance of the prior probability assumptions present in Bayesian models. However, particular attention is paid in this point because age determination techniques must be also tested with other methods as for instance gyrochronology, asteroseismology, chemical clocks, etc. In this paper, we adopted the ages estimated using the \qq\ since it is shown that this method gives reliable ages as the Bayesian inference. Therefore, the \qq\ ages will be used in the figures of the next sections. All our age, mass, and radius results using both methods are summarized in Table \ref{appendix_tab:2}.

\begin{figure*}
 \includegraphics[scale=0.5]{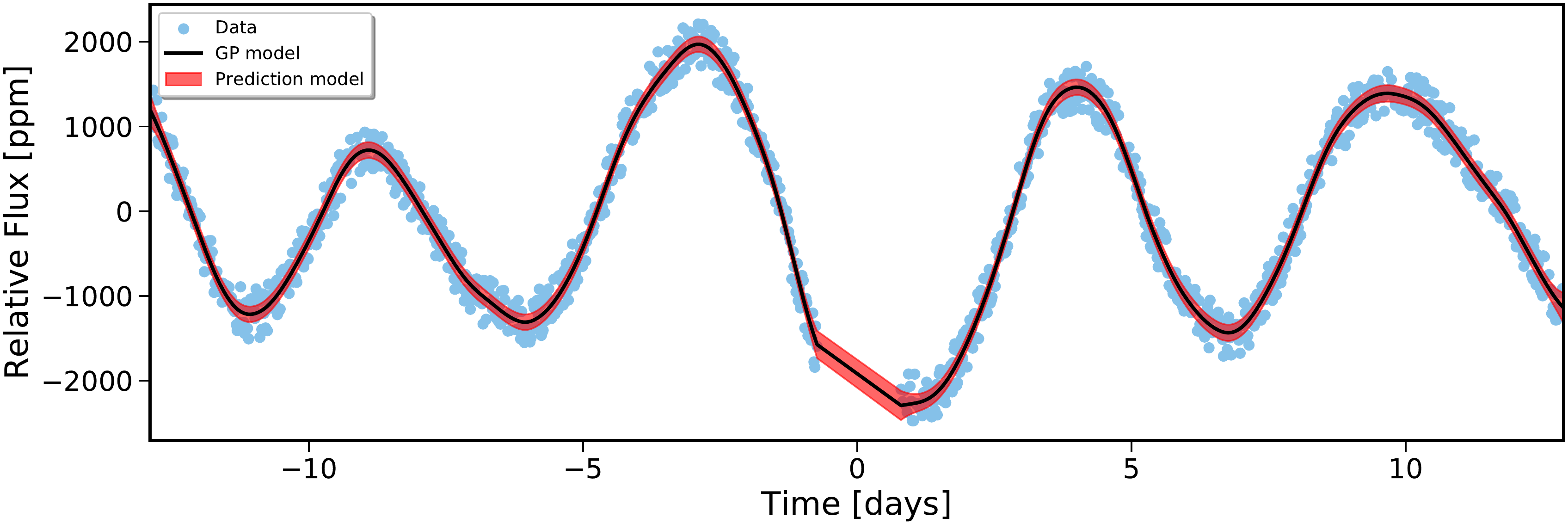}
 \centering
 \caption{TESS light curve (filled circles) of the new solar proxy HIP 17936 observed in one sector. The black line represents the GP model, while the red shaded region its 2$\sigma$ rotation prediction model.}
 \label{fig:RP_GP_3277015039888106496}
\end{figure*} 

\section{Stellar Rotation with Kepler \& \tess}
It is well-known that late-type stars inherit part of the original molecular cloud angular momentum as they are born. Therefore, a large spread in their initial rotational velocities is observed among young open cluster and stellar associations \citep{Bouvier:1997A&A...326.1023B}. As the stars arrive at the main sequence, it is expected that magnetized stellar winds, powered by stellar dynamo, drive the angular momentum evolution throughout their evolutionary history, gradually forgetting the initial rotational conditions. Therefore, after a given age, late-type stars tend to converge into well-behaved rotational sequence as a function of mass and age, enabling the calibration of empirical rotation-age-mass relations \citep{Skumanich:1972ApJ...171..565S, Barnes:2003ApJ...586..464B}. This age-dating technique is known as gyrochronology and establish a precise rotational clock where stellar ages are estimated from rotational period measurements \citep{Barnes:2003ApJ...586..464B, Meibom:2011ApJ...733..115M, Meibom:2015Natur.517..589M}. 

In the last decade, a new era for astronomy began with the successful \kepler\ \citep{Kepler:2010Sci...327..977B}, K2 \citep{K2:2014PASP..126..398H}, and \tess\ \citep{TESS:2015JATIS...1a4003R} missions. In this paper, we take advantage of the large public database of these surveys to measure rotation periods (\prot) for our sample. We found 31 precise light curves (one in \kepler, two in K2, and 28 in TESS) where several of them belong to the short and long cadence (e.g., 2-minute and 30-minute cadence observation in TESS). To extract TESS and Kepler light curves, we repeated the same procedure adopted in \citet{Diego:2020MNRAS.495L..61L}. The light curves were obtained from target pixel files using pixel level decorrelation technique through the \textit{lightkurve}\footnote{\url{https://docs.lightkurve.org/index.html}} python package \citep{Lightkurve:2018ascl.soft12013L}. For each target, we remove surrounding pixels eventually contaminated by nearby stars. The resulting light curves are cleaned from outliers beyond $\pm$3$\sigma$ and in some cases binned in steps of 0.5 h to enhance the signal-to-noise ratio and also mitigate short-term variability (e.g., oscillations, spacecraft pointing jitter).

\prot\ were initially measured through Generalized Lomb-Scargle analysis. Since most of our stars shows moderate to low level of activity, we restricted our search for rotational periods within a reasonable window between 1 and 50 days. Detected rotation periods are defined by signals in the periodograms with false alarm probability bellow 1\%. In the cases where aliases of the strongest detection are also present and statistically significant, we choose to report the secondary detection together with the strongest one. For a sanity test, the \prot\ were also estimated through Gaussian Process (GP) that uses a kernel developed from a mixture of two harmonic oscillators \cite[for a complete description see][]{exoplanet:exoplanet}. Figure \ref{fig:RP_GP_3277015039888106496} shows the reduced light curve of HIP 17936 with the GP model plotted in black solid line and its 2$\sigma$ model prediction in red shaded region. The \prot\ estimated with both methods are in a very good agreement and we use the median of them as the adopted rotational period value (see Table \ref{appendix_tab:1}). 

\begin{figure}
 \includegraphics[scale=0.53]{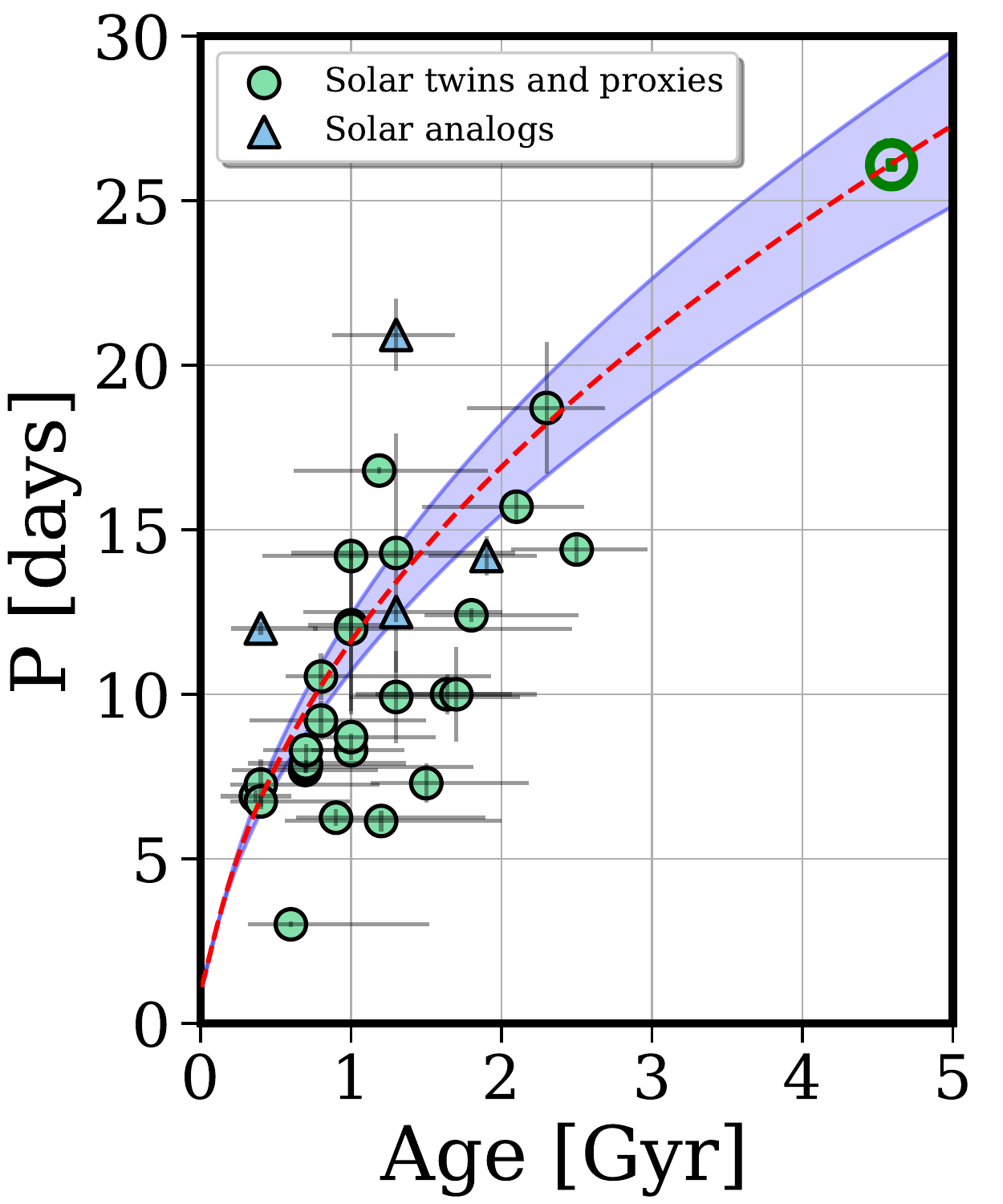}
 \centering
 \caption{Age-Rotation diagram for our solar twins and proxies using model predictions from \citet{Barnes:2010ApJ...722..222B} calibrated to the Sun (green solar standard symbol) and with variations of $\pm$0.05 M$_{\odot}$ (shaded region). The red dashed line represents the rotational evolution of the Sun.}
 \label{fig:Age_rotation}
\end{figure}

In Figure \ref{fig:Age_rotation} are displayed the rotational evolution of our sample of solar proxies (green circles), solar analogs (blue triangles) and the Sun (green solar standard symbol). The shaded region represents the rotational evolution model with solar metallicity and mass within $\pm$5\% of the Sun. The model is based on the \textit{period evolution equation} established by \citet{Barnes:2010ApJ...722..222B} (see Equation (32) therein) that includes the convective turnover timescale \citep[Table 1,][]{Barnes:2010ApJ...721..675B}. We found a good agreement between the isochronal \qq\ ages and the rotational evolution model, almost within the uncertainties.

\section{Chromospheric activity}
\begin{figure}
 \includegraphics[scale=0.53]{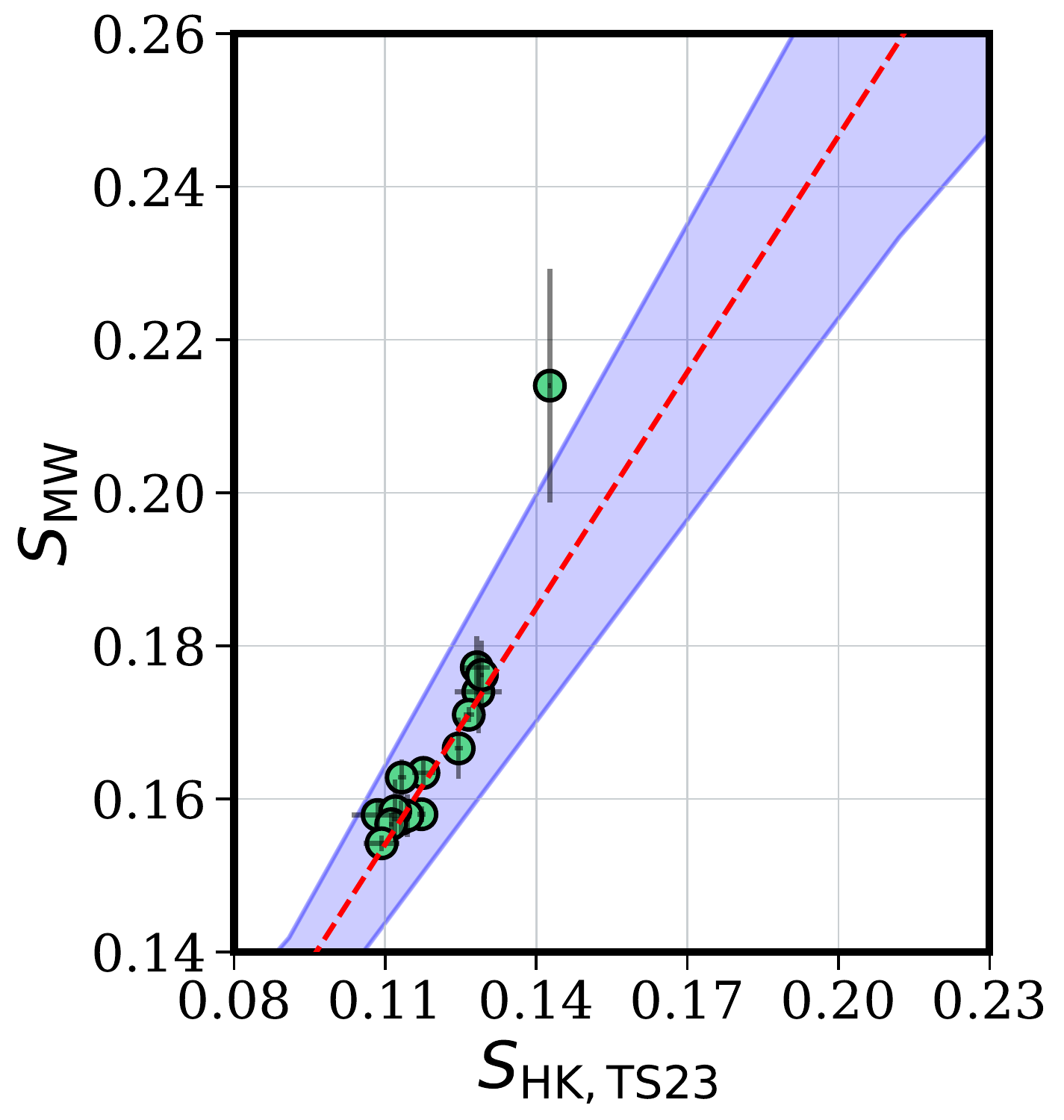}
 \centering
 \caption{Mount Wilson system calibration using $S_{\rm{HK,TS23}}$ versus $S_{\rm{HK,MW}}$. The red dashed line and the blue shaded region represent the linear fit (considering the errors in both axes) and the 95\% confidence interval, respectively.}
 \label{fig:S_index_calibration}
\end{figure}

Thanks to the very good performance of the TS23 spectrograph in the blue part of the spectra, we estimated the activity indices for our sample by measuring the \ion{Ca}{II} H\&K emission line fluxes (3933.664 \AA\ and 3968.470 \AA). The normalization of the spectral region bracketing the \ion{Ca}{II} lines demands a different normalization procedure. In order to ensure the overall consistency of activity measurements, for each star, we performed a differential normalization procedure of the \textit{echelle} spectral orders that surround the \ion{Ca}{II} lines. As a template to guide the normalization procedure of a given star, we build a high SNR master spectrum from the large HARPS ($R$ = 115000) time series and thus degraded the resolving power to match with TS23 observations ($R$ = 60000). In the cases where no HARPS observations were performed for a given TS23 solar twin candidate, we choose another HARPS star with similar \teff\ and \feh\, as a template. The $S_{\rm{HK}}$ index was calculated following the prescription given in \cite{Wright:2004ApJS..152..261W}. In order to perform a reliable calibration of our $S_{\rm{HK, TS23}}$ indices into the Mount Wilson system (MW), we first selected a subsample of 10 stars whose $S_{\rm MW}$ are very well estimated by \citet{Diego:2018A&A...619A..73L} and then complemented with 5 new solar-type stars found in the European Southern Observatory (ESO) archive\footnote{\url{http://archive.eso.org/wdb/wdb/adp/phase3_spectral/form?phase3_collection=HARPS}}  also with common observations between HARPS and TS23. This sample of 15 stars is distributed between active and inactive regimes (Table \ref{appendix_tab:3}). As a result, we obtain the following calibration equation:
\be
\label{eq:calib_index}
S_{\rm{MW}} = 0.038(\pm0.015) + 1.048(\pm0.125)\times S_{\rm{HK, TS23}}, 
\ee
where the typical standard deviation of the linear fit is 0.0073 (see Figure \ref{fig:S_index_calibration}), comparable with calibrations carried out using spectrographs of higher resolving power and stability than TS23 (e.g, HARPS). The converted $S_{\rm{MW}}$ values and their respective errors taking into account photometric and repeatability measurement errors are given in Table \ref{appendix_tab:1} for the \inti\ sample. However, we excluded from our sample stars with $S_{\rm{MW}}$ values estimated in a single patch epoch.

Activity levels were estimated using $\log R'_{\rm{HK}} (T_{\rm eff})$ index following the procedure given by \cite{Diego:2018A&A...619A..73L}. We emphasize that $\log R'_{\rm{HK}} (T_{\rm eff})$ should not be confused with the usual $\log R'_{\rm{HK}}$ based on photometric colors \citep{Noyes:1984ApJ...279..763N}. For the most active stars, the difference between both indices is negligible, however substantial differences arise after $\log R'_{\rm{HK}}\sim-4.8$ towards the lowest activity levels. Besides, the updated \ion{Ca}{II} index $\log R'_{\rm{HK}} (T_{\rm eff})$ shows improved activity-age correlation for inactive stars. To build this index, \cite{Diego:2018A&A...619A..73L} removed the photospheric contribution of the $R'_{\rm{HK}}$ by using an improved photospheric correction as a function of \teff\ (Equation (7) therein) instead of the standard photometric color $(B-V)$ \citep{Wright:2004ApJS..152..261W}. The uncertainties are estimated through random samples from a Gaussian distribution that takes into account the $\sigma(S_{\rm{HK,MW}})$. As a result, we obtained updated chromospheric indices $\log R'_{\rm{HK}}$(\teff) for our sample and they can be also found in Table \ref{appendix_tab:1}.

\begin{figure}
 \includegraphics[width=\columnwidth]{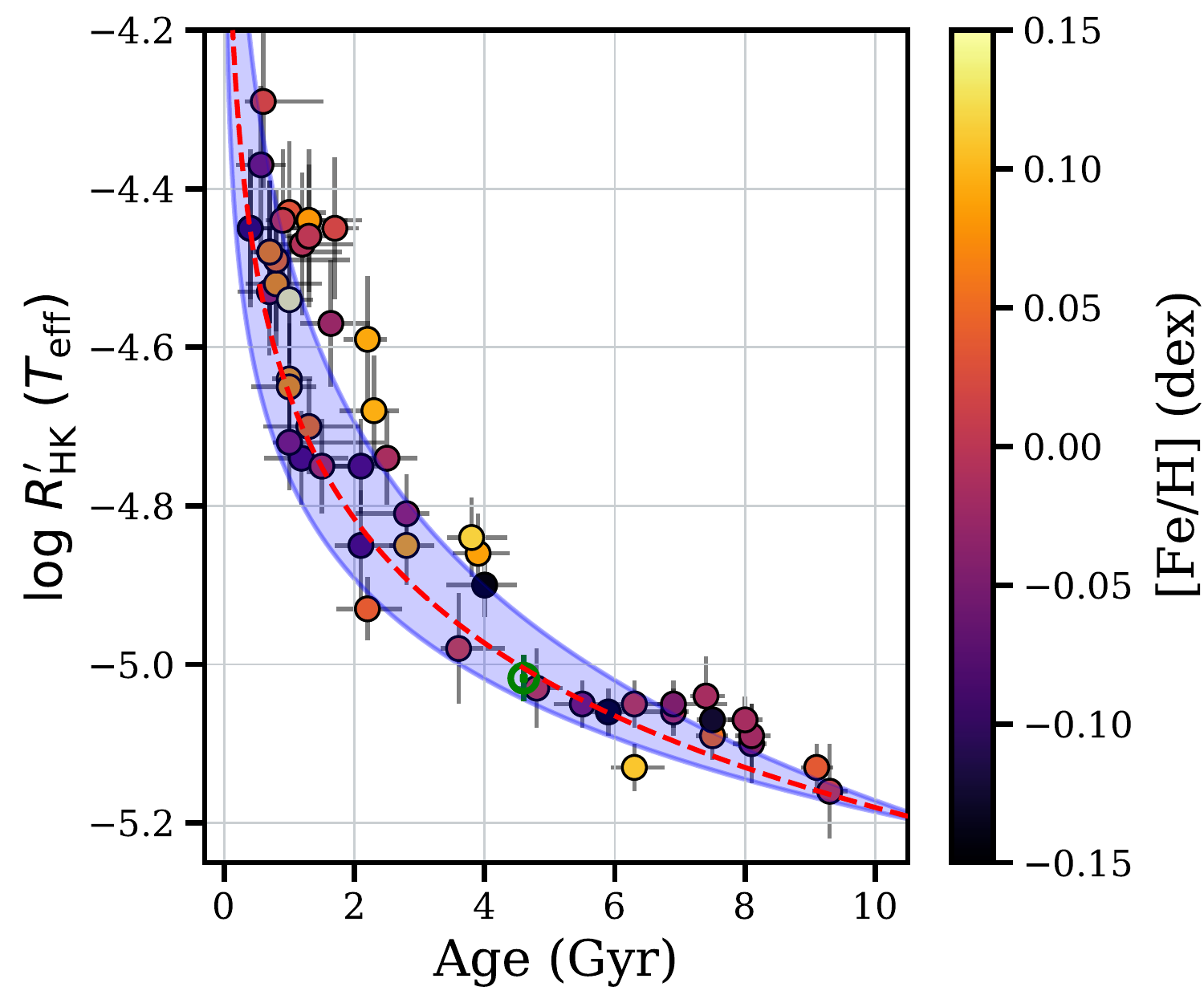}
 \centering
 \caption{Activity-age diagram for our solar twins and solar proxies. The red dashed line represents the activity-age relation ($R'_{\rm{HK}}$(\teff) $\propto \rm{Age}^{-0.52}$) and the shaded region its $2\sigma$ activity prediction band found by \citet{Diego:2018A&A...619A..73L}. The Sun is represented by the green solar standard symbol, while the colormap shows the metallicity distribution.}
 \label{fig:activity_age_inti}
\end{figure}

In Figure \ref{fig:activity_age_inti} are shown the chromospheric indices versus the ages for our solar proxy sample. The activity-age relation found by \citet{Diego:2018A&A...619A..73L} is also plotted as red dashed lines with its 2$\sigma$ activity variability prediction band (shaded region). We can clearly see that our new sample of solar twins/proxies also follows this correlation, thereby favouring the chromospheric activity as a useful clock even for stars older than the Sun. However, notice that the activity-age relation in \citet{Diego:2018A&A...619A..73L} was derived using high resolving power ($R$ = 115000) time-series of \ion{Ca}{II} H\&K measurements. Besides, the typical \feh\ values of their sample are more narrowed around the solar metallicity ($\pm$0.05 dex of the solar value) in comparison to our sample. The latter explains the presence of some outliers in our activity-age diagram. The solar analog stars and solar-type stars are not included in Figure \ref{fig:activity_age_inti} as their mass and \feh\ regime are different to the solar proxies.

\section{Summary and conclusions}
Thanks to the \textit{Gaia} mission, we found a large sample of solar twins candidates through constraints on color (using Tycho, 2MASS and \gaiaedr\ catalogs) and absolute magnitude (employing \gaiaedr\ parallaxes). As our sample is within 100 pc from Earth, reddening corrections are negligible. The definitive color constraints used for our solar twin hunting program are shown in Table \ref{tab:1} and were established following the spectroscopic solar twin definition given by \cite{Ramirez:2014A&A...572A..48R}. However, this definition does not consider the evolutionary state of the star, thus introducing a slight bias in the mass distribution in the sample of known solar twins \citep[see][]{Ramirez:2014A&A...572A..48R, Spina:2018MNRAS.474.2580S}. Although the selection criteria of \citet{Ramirez:2009A&A...508L..17R} is useful for obtaining precise chemical abundances differentially to the Sun (due to a narrow range in stellar parameters relative to the Sun), that criteria hampers studies dedicated to understand the rotational and magnetic evolution of the Sun. To address this issue for future works, we propose a new class of star like the Sun: \textit{solar proxy}, whose definition is based not only on stellar parameters constraints, but also on its evolutionary track during the main sequence. In this new definition, the metallicity and the mass define whether a star is a solar proxy or not. These parameters are constrained to be within $\pm$0.15 dex and 5\% to the solar values, thereby assuring that the star follows a similar evolution as the Sun. The \logg\ is assumed to be from $\sim$ 4.1 to 4.6 dex since this constraint is used only to verify if the star is on the main sequence. As is shown in Figure \ref{fig:McDonald_nst}, the \teff\ of solar proxies can take values ranging from $\sim$5310$-$6050 K, however, for precise abundances, it is recommended to work with stars with \teff\ within 150-200 K of the solar value in order to avoid 3D and non-LTE effects. Note that the solar proxy limits are not defined by an irregular polygon region, but instead by evolutionary tracks.

Applying all the definitions discussed above, we identified 70 solar proxies, 46 solar analogs and 13 solar-like stars. Their stellar parameters were estimated through the differential analysis and the spectroscopic equilibrium technique. As a result, we obtained a high internal precision ($\sigma$(\teff) = 15 K, $\sigma$(\logg) = 0.03 dex, $\sigma$(\feh) = 0.01 dex, and $\sigma$(\vmic) = 0.03 km s$^{-1}$). We also search the \textit{close solar twin} within our \inti\ sample by narrowing down the mass, \teff, \logg, and \feh\ values to 0.03 \sm, 50 K, 0.05 dex, and 0.05 dex relative to the solar values, respectively. We found 9 potential candidates that meet these rigorous criteria. However, further studies should be performed to confirm these stars as close solar twins (e.g., chemical composition, Li abundances, etc).

Isochronal ages were estimated trough the Yonsei-Yale ischrones models \citep{Yi:2001ApJS..136..417Y} and employing two algorithms that use spectroscopic stellar parameters and precise \gaiaedr\ parallaxes as input parameters. The ages and masses show a good agreement between the methods within the uncertainties. We also estimated the trigonometric gravity to compare with the spectroscopic gravity and we found relatively good agreement between them, thus validating our age results. However, this is not the case for some binary stars, and it is necessary to use other age determinations such as gyrochronology, asteroseismology, chromospheric-age relations, etc., in order to evaluate the reliability of the results. We also determine a precise Mount Wilson system calibration for the activity indices ($S_{\rm{HK, TS23}}$) taken with the TS23 spectrograph at McDonald Observatory (see Eq. \ref{eq:calib_index} and Figure \ref{fig:S_index_calibration}). With this new calibration, we obtained improved chromospheric indices ($\log R'_{\rm{HK}}$(\teff); \citealp{Diego:2018A&A...619A..73L}). Our new sample of solar twins/proxies also follow the activity-age correlation, thereby reinforcing the scenario where stars older than the Sun continue to decrease their chromospheric activity (see Figure \ref{fig:activity_age_inti}). Rotational periods were estimated using precise TESS, \kepler, and K2 light curves after applying the Generalized Lomb-Scargle and Gaussian Process methods. 

In this work, we provide to the community precise stellar parameters, ages, chromospheric indices and rotational periods (albeit we were not able to detect rotational periods in stars older than the Sun; Figure \ref{fig:Age_rotation}). Finally, the \inti\ survey is ideal for exoplanet searches around stars like the Sun (e.g., \citealp{Bedell:2015A&A...581A..34B}), in the quest for Solar System analogs.

\section*{Acknowledgments}
\textit{Research funding agencies:} J.Y.G. acknowledges the support from CNPq. D.L.O. and J.M. thank the support from FAPESP (2016/20667-8; 2018/04055-8). M.F. and P.M. acknowledge the financial support from CONICET in the forms of Doctoral and Post-Doctoral Fellowships. \\

\textit{Dedication:} This work is dedicated to the memory of Giusa Cayrel de Strobel, who had tremendous influence and contribution to the stellar astrophysics that we know today. This paper is purely inspired in Cayrel de Strobel's work who initiated the search for \textit{real solar twins} in 1981. This work is also dedicated to Peru's Bicentennial Generation (\textit{Generaci\'on del Bicentenario}) and overall to Brian Pintado S\'anches and Inti Sotelo Camargo, who lost their lives defending the Peruvian democracy on November 14th, 2020. \\

\textit{Special acknowledgments:} We would like to thank Chris Sneden and Ivan Ram\'irez for providing us their useful notes of observing and reducing TS23 spectra, that helped us to develop the \textit{TS23\_reduc} pipeline. We thank Bruno Quint, Tina Armond, Raquel Santiago, and Nat\'alia Amarinho for kindly supporting us during our observations with the SOAR telescope. We also thank Candace Gray from Apache Point Observatory for kindly providing the ARCES twilight solar spectra. \\

\textit{Facilities:} \textbf{McDonald Observatory:} Harlam J. Smith 2.7-meter Telescope,  TS23 spectrograph. \textbf{Apache Point Observatory:} Astrophysical Research Consortium 3.5-meter Telescope, ARCES spectrograph. \textbf{SOAR Telescope at Cerro Pachon:} The Southern Astrophysical Research (SOAR) 4.1-meter Telescope, Goodman High Throughput Spectrograph. \\

\textit{Softwares:} \textbf{numpy} \citep{van_der_Walt:2011CSE....13b..22V}, \textbf{matplotlib} \citep{Hunter:4160265}, \textbf{pandas} \citep{mckinney-proc-scipy-2010}, \textbf{astroquery} \citep{Ginsburg:2019AJ....157...98G}, \textbf{IRAF} \citep{Tody:1986SPIE..627..733T}, \textbf{iSpec} \citep{Blanco:2014A&A...569A.111B, Blanco:2019MNRAS.486.2075B}, \textbf{Kapteyn Package} \citep{KapteynPackage}, \textbf{MOOG} \citep{Sneden:1973PhDT.......180S}, \textbf{q$^{\textbf{2}}$} \citep{Ramirez:2014A&A...572A..48R}, \textbf{CERES} \citep{CERES:2017PASP..129c4002B}, \textbf{SP\_ACE} \citep{SPACE:2016A&A...587A...2B}. \\

\section*{Data Availability}
The \inti\ survey is available through MNRAS in its online supplementary material, and its spectra can be shared on request to the corresponding author.


\bibliographystyle{mnras}
\bibliography{references}

\appendix
\section{Tables}
\begin{landscape}
\begin{table}
    \centering
    \caption{Stellar parameters for the \inti\ survey. This table is available in its entirety in machine readable format at the CDS.}
    \label{appendix_tab:1}
    \begin{tabular}{ccccccccccc}
        \hline
        \hline
        \textit{Gaia} EDR3  &   Identifier   &     \teff\ (K)   &    \logg\ (dex)   & \logg$_{\rm{Trigonometric}}$ (dex)   &    \feh\ (dex)     & \vmic\ (km s$^{-1}$) & $S_{\rm{MW}}$       & $\log R'_{\rm{HK}}$(\teff)  &     \prot      & Remarks \\
        \hline                                                                                                                                                                                                                             
        2117267079903573504 & TYC3130-2191-1 &    5722 $\pm$ 8  & 4.350 $\pm$ 0.025 &          4.380 $\pm$ 0.023           & -0.040 $\pm$ 0.007 &    1.03 $\pm$ 0.02   &  0.161 $\pm$ 0.006  &        -5.10 $\pm$ 0.05     &                & 1       \\
        297548019938221056  & HIP8522        &    5727 $\pm$ 13 & 4.520 $\pm$ 0.026 &          4.543 $\pm$ 0.023           & -0.017 $\pm$ 0.010 &    1.11 $\pm$ 0.03   &  0.324 $\pm$ 0.003  &        -4.53 $\pm$ 0.08     & 7.7  $\pm$ 0.1 & 1       \\
        4529634417760144896 & HD168069       &    5871 $\pm$ 13 & 4.490 $\pm$ 0.027 &          4.466 $\pm$ 0.023           &  0.038 $\pm$ 0.009 &    1.05 $\pm$ 0.02   &  0.178 $\pm$ 0.003  &        -4.93 $\pm$ 0.04     &                & 1       \\
        1761984258439478784 & HIP103025      &    5827 $\pm$ 12 & 4.380 $\pm$ 0.035 &          4.339 $\pm$ 0.023           &  0.013 $\pm$ 0.010 &    1.14 $\pm$ 0.02   &  0.161 $\pm$ 0.003  &        -5.05 $\pm$ 0.03     &                & 1       \\
        990844114760515840  & HD49425        &    5779 $\pm$ 9  & 4.430 $\pm$ 0.026 &          4.429 $\pm$ 0.023           & -0.026 $\pm$ 0.008 &    1.11 $\pm$ 0.02   &                     &                             &                & 1       \\
        272026980672350336  & HIP20722       &    5846 $\pm$ 8  & 4.480 $\pm$ 0.028 &          4.493 $\pm$ 0.023           &  0.092 $\pm$ 0.008 &    1.10 $\pm$ 0.02   &  0.257 $\pm$ 0.027  &        -4.64 $\pm$ 0.10     & 8.3  $\pm$ 0.3 & 1       \\
        740735868326858368  & HIP49580       &    5801 $\pm$ 10 & 4.470 $\pm$ 0.033 &          4.487 $\pm$ 0.023           & -0.023 $\pm$ 0.009 &    1.03 $\pm$ 0.02   &  0.210 $\pm$ 0.003  &        -4.81 $\pm$ 0.05     &                & 1       \\
        1508557582834745088 & HIP69709       &    5856 $\pm$ 11 & 4.320 $\pm$ 0.032 &          4.322 $\pm$ 0.023           & -0.007 $\pm$ 0.009 &    1.14 $\pm$ 0.02   &  0.159 $\pm$ 0.003  &        -5.06 $\pm$ 0.03     &                & 1       \\
        3557158782894062080 & HIP52649       &    5811 $\pm$ 11 & 4.340 $\pm$ 0.028 &          4.336 $\pm$ 0.023           & -0.046 $\pm$ 0.009 &    1.10 $\pm$ 0.02   &  0.163 $\pm$ 0.003  &        -5.05 $\pm$ 0.03     &                & 1       \\
        1260316891261098752 & HIP69230       &    5734 $\pm$ 11 & 4.490 $\pm$ 0.031 &          4.521 $\pm$ 0.023           & -0.078 $\pm$ 0.009 &    1.03 $\pm$ 0.03   &  0.238 $\pm$ 0.003  &        -4.74 $\pm$ 0.06     & 16.8 $\pm$ 0.1 & 1       \\
        3915429733361853952 & HIP55820       &    5672 $\pm$ 8  & 4.440 $\pm$ 0.027 &          4.463 $\pm$ 0.023           & -0.044 $\pm$ 0.007 &    0.98 $\pm$ 0.02   &  0.171 $\pm$ 0.003  &        -5.05 $\pm$ 0.03     &                & 1       \\
        2136741591197747712 & HD184424       &    5827 $\pm$ 10 & 4.470 $\pm$ 0.034 &          4.497 $\pm$ 0.023           &  0.083 $\pm$ 0.009 &    1.19 $\pm$ 0.02   &  0.308 $\pm$ 0.009  &        -4.52 $\pm$ 0.08     & 9.2  $\pm$ 0.6 & 1       \\
        345765551891443968  & HIP9172        &    5794 $\pm$ 16 & 4.560 $\pm$ 0.030 &          4.533 $\pm$ 0.023           &  0.031 $\pm$ 0.013 &    1.24 $\pm$ 0.03   &  0.367 $\pm$ 0.009  &        -4.43 $\pm$ 0.09     & 8.7  $\pm$ 0.1 & 1       \\
        637329067477530368  & HIP46076       &    5766 $\pm$ 11 & 4.470 $\pm$ 0.030 &          4.509 $\pm$ 0.023           & -0.076 $\pm$ 0.010 &    1.05 $\pm$ 0.03   &  0.229 $\pm$ 0.005  &        -4.75 $\pm$ 0.06     & 15.7 $\pm$ 0.5 & 1       \\
        3203761005400205824 & HIP20218       &    5755 $\pm$ 9  & 4.450 $\pm$ 0.030 &          4.500 $\pm$ 0.023           & -0.015 $\pm$ 0.009 &    1.09 $\pm$ 0.02   &  0.234 $\pm$ 0.003  &        -4.74 $\pm$ 0.06     & 14.4 $\pm$ 0.5 & 1       \\
        1323563308352525440 & HIP79068       &    5874 $\pm$ 7  & 4.480 $\pm$ 0.022 &          4.492 $\pm$ 0.023           &  0.084 $\pm$ 0.007 &    1.09 $\pm$ 0.02   &  0.252 $\pm$ 0.016  &        -4.65 $\pm$ 0.08     & 14.2 $\pm$ 0.1 & 1       \\
        882410278028835584  & HIP38647       &    5767 $\pm$ 15 & 4.510 $\pm$ 0.026 &          4.517 $\pm$ 0.023           & -0.006 $\pm$ 0.012 &    1.11 $\pm$ 0.03   &  0.231 $\pm$ 0.004  &        -4.75 $\pm$ 0.06     & 7.3  $\pm$ 0.6 & 1       \\
        291701396922348672  & HIP7244        &    5766 $\pm$ 10 & 4.490 $\pm$ 0.024 &          4.519 $\pm$ 0.023           & -0.028 $\pm$ 0.009 &    1.13 $\pm$ 0.02   &  0.298 $\pm$ 0.012  &        -4.57 $\pm$ 0.08     & 10.0 $\pm$ 0.6 & 1       \\
        333106054183616000  & HIP10321       &    5725 $\pm$ 10 & 4.510 $\pm$ 0.022 &          4.523 $\pm$ 0.023           & -0.020 $\pm$ 0.007 &    0.97 $\pm$ 0.02   &  0.347 $\pm$ 0.006  &        -4.48 $\pm$ 0.09     & 7.8  $\pm$ 0.2 & 1       \\
        4511284599484268544 & HD229450A      &    5739 $\pm$ 8  & 4.335 $\pm$ 0.028 &          4.367 $\pm$ 0.023           &  0.109 $\pm$ 0.009 &    1.04 $\pm$ 0.02   &  0.156 $\pm$ 0.003  &        -5.13 $\pm$ 0.03     &                & 1       \\
        2869747927140286208 & HD221103       &    5879 $\pm$ 9  & 4.445 $\pm$ 0.028 &          4.426 $\pm$ 0.023           &  0.097 $\pm$ 0.008 &    1.14 $\pm$ 0.02   &  0.194 $\pm$ 0.003  &        -4.85 $\pm$ 0.05     &                & 1       \\
        2163433717746084352 & TYC3588-6756-1 &    5812 $\pm$ 23 & 4.530 $\pm$ 0.038 &          4.513 $\pm$ 0.023           &  0.015 $\pm$ 0.017 &    1.45 $\pm$ 0.04   &  0.464 $\pm$ 0.010  &        -4.29 $\pm$ 0.11     & 3.0  $\pm$ 0.1 & 1       \\
        4534307273451743744 & HD342979       &    5667 $\pm$ 10 & 4.470 $\pm$ 0.034 &          4.524 $\pm$ 0.023           &  0.060 $\pm$ 0.009 &    0.99 $\pm$ 0.03   &  0.259 $\pm$ 0.006  &        -4.70 $\pm$ 0.06     & 14.3 $\pm$ 3.6 & 1       \\
        1855968480863250176 & HIP101893      &    5729 $\pm$ 11 & 4.510 $\pm$ 0.026 &          4.514 $\pm$ 0.023           &  0.058 $\pm$ 0.009 &    1.17 $\pm$ 0.02   &  0.345 $\pm$ 0.004  &        -4.49 $\pm$ 0.09     & 10.5 $\pm$ 0.7 & 1       \\
        1909471786118933120 & HD213244       &    5689 $\pm$ 9  & 4.465 $\pm$ 0.031 &          4.533 $\pm$ 0.023           & -0.006 $\pm$ 0.008 &    1.09 $\pm$ 0.03   &  0.362 $\pm$ 0.008  &        -4.47 $\pm$ 0.09     & 6.2  $\pm$ 0.3 & 1       \\
        2196984318618671360 & TYC4250-1364-1 &    5747 $\pm$ 8  & 4.480 $\pm$ 0.033 &          4.493 $\pm$ 0.023           &  0.092 $\pm$ 0.008 &    1.12 $\pm$ 0.02   &  0.291 $\pm$ 0.007  &        -4.59 $\pm$ 0.08     &                & 1       \\
        1792706331305783808 & TYC1678-109-1  &    5769 $\pm$ 11 & 4.460 $\pm$ 0.033 &          4.499 $\pm$ 0.023           & -0.005 $\pm$ 0.010 &    1.18 $\pm$ 0.03   &                     &                             &                & 1       \\
        1962597885872344704 & TYC3210-873-1  &    5742 $\pm$ 11 & 4.500 $\pm$ 0.030 &          4.508 $\pm$ 0.023           &  0.080 $\pm$ 0.010 &    1.16 $\pm$ 0.03   &  0.370 $\pm$ 0.004  &        -4.44 $\pm$ 0.09     & 9.9  $\pm$ 1.4 & 1       \\
        2003130042007448704 & HD235929       &    5668 $\pm$ 23 & 4.530 $\pm$ 0.040 &          4.553 $\pm$ 0.023           &  0.007 $\pm$ 0.015 &    1.17 $\pm$ 0.04   &  0.385 $\pm$ 0.003  &        -4.44 $\pm$ 0.09     & 6.3  $\pm$ 0.3 & 1       \\
        1796825582900925056 & TYC2211-2123-1 &    5775 $\pm$ 9  & 4.415 $\pm$ 0.026 &          4.447 $\pm$ 0.023           &  0.087 $\pm$ 0.008 &    1.06 $\pm$ 0.02   &  0.199 $\pm$ 0.005  &        -4.86 $\pm$ 0.05     &                & 1       \\
        1957736635726467840 & TYC3208-1984-1 &    5675 $\pm$ 8  & 4.433 $\pm$ 0.027 &          4.512 $\pm$ 0.023           &  0.096 $\pm$ 0.007 &    1.08 $\pm$ 0.02   &  0.265 $\pm$ 0.005  &        -4.68 $\pm$ 0.07     & 18.7 $\pm$ 2.0 & 1       \\
        1989930095677359616 & HD235957       &    5707 $\pm$ 10 & 4.320 $\pm$ 0.029 &          4.375 $\pm$ 0.023           & -0.020 $\pm$ 0.008 &    1.08 $\pm$ 0.02   &  0.163 $\pm$ 0.004  &        -5.09 $\pm$ 0.04     &                & 1       \\
        225038148667264384  & HD22875        &    5732 $\pm$ 9  & 4.485 $\pm$ 0.034 &          4.513 $\pm$ 0.023           &  0.018 $\pm$ 0.009 &    1.21 $\pm$ 0.03   &  0.366 $\pm$ 0.004  &        -4.45 $\pm$ 0.09     & 10.0 $\pm$ 1.4 & 1       \\
        1807690750662760576 & HD190340       &    5872 $\pm$ 10 & 4.390 $\pm$ 0.029 &          4.403 $\pm$ 0.023           &  0.109 $\pm$ 0.009 &    1.12 $\pm$ 0.02   &  0.228 $\pm$ 0.003  &        -4.72 $\pm$ 0.06     &                & 1\dag   \\
        2592050460064616832 & HD9201         &    5786 $\pm$ 11 & 4.440 $\pm$ 0.028 &          4.466 $\pm$ 0.023           &  0.035 $\pm$ 0.009 &    1.06 $\pm$ 0.02   &                     &                             &                & 1       \\
        132371961511017856  & HIP11253       &    5812 $\pm$ 15 & 4.430 $\pm$ 0.041 &          4.328 $\pm$ 0.023           & -0.006 $\pm$ 0.012 &    1.17 $\pm$ 0.03   &  0.180 $\pm$ 0.003  &        -4.94 $\pm$ 0.04     &                & 1\dag   \\
        1605622057017876992 & HIP71989       &    5890 $\pm$ 16 & 4.540 $\pm$ 0.024 &          4.527 $\pm$ 0.023           & -0.016 $\pm$ 0.010 &    1.00 $\pm$ 0.03   &  0.298 $\pm$ 0.006  &        -4.53 $\pm$ 0.08     & 6.9  $\pm$ 0.2 & 1\dag   \\
        2257140386478701568 & HIP91210       &    5869 $\pm$ 8  & 4.480 $\pm$ 0.029 &          4.488 $\pm$ 0.023           &  0.029 $\pm$ 0.008 &    1.10 $\pm$ 0.02   &  0.244 $\pm$ 0.003  &        -4.67 $\pm$ 0.06     &                & 1\dag   \\
        3827020717791837056 & HIP47312       &    5884 $\pm$ 9  & 4.430 $\pm$ 0.025 &          4.410 $\pm$ 0.023           & -0.010 $\pm$ 0.007 &    1.11 $\pm$ 0.02   &  0.172 $\pm$ 0.004  &        -4.96 $\pm$ 0.04     &                & 1\dag   \\
        855523714036230016  & HIP52278       &    5727 $\pm$ 9  & 4.510 $\pm$ 0.019 &          4.505 $\pm$ 0.023           & -0.079 $\pm$ 0.006 &    1.02 $\pm$ 0.02   &  0.208 $\pm$ 0.014  &        -4.85 $\pm$ 0.07     &                & 1\dag   \\
        3574377616021489024 & HD105590A      &    5759 $\pm$ 8  & 4.470 $\pm$ 0.026 &          4.439 $\pm$ 0.023           &  0.010 $\pm$ 0.008 &    1.03 $\pm$ 0.02   &  0.162 $\pm$ 0.003  &        -5.08 $\pm$ 0.03     &                & 1\dag   \\
        1610658885425227264 & HIP71440       &    5782 $\pm$ 20 & 4.510 $\pm$ 0.036 &          4.263 $\pm$ 0.025           & -0.046 $\pm$ 0.015 &    1.17 $\pm$ 0.03   &  0.319 $\pm$ 0.004  &        -4.52 $\pm$ 0.08     & 6.2  $\pm$ 0.3 & 1\dag   \\
        \hline
        \hline
    \end{tabular}
    \begin{tablenotes}
	\item \textbf{Notes.}(1) Solar twins, (2) Solar proxies, (3) Solar analogs, (4) Solar-type stars, (\dag) Binary star.
	\end{tablenotes}
\end{table}
\end{landscape}

\begin{landscape}
\begin{table}
    \centering
    \caption{Fundamental parameters for the \inti\ survey. This table is available in its entirety in machine readable format at the CDS.}
    \label{appendix_tab:2}
    \begin{tabular}{ccccccccc}
        \hline
        \hline
        \textit{Gaia} EDR3 & Identifier     &       Age (Gyr)        &        Mass (\sm)        &      Radius (\sr)        &        Age (Gyr)      &         Mass (\sm)      &       Radius (\sr)      & Remarks \\
                            &                & (\qq \& \logg\ \& plx) & (\qq \& \logg\ \& plx)   & (\qq \& \logg\ \& plx)   &       (Bayesian)      &       (Bayesian)        &      (Bayesian)         &         \\
        \hline                                                                                                                                                                                     
        2117267079903573504 & TYC3130-2191-1 &  8.10  (+0.52)(-0.29)  &  0.960 (+0.014)(-0.007)  &  1.060 (+0.014)(-0.008)  &   8.60 (+0.40)(-0.60) &  0.940 (+0.010)(-0.008) & 1.045 (+0.010)(-0.013)  & 1       \\
        297548019938221056  & HIP8522        &  0.70  (+0.97)(-0.49)  &  0.999 (+0.004)(-0.004)  &  0.902 (+0.004)(-0.004)  &   0.60 (+0.50)(-0.40) &  0.998 (+0.008)(-0.008) & 0.893 (+0.008)(-0.010)  & 1       \\
        4529634417760144896 & HD168069       &  2.20  (+1.01)(-0.48)  &  1.050 (+0.008)(-0.008)  &  1.000 (+0.010)(-0.009)  &   2.20 (+0.50)(-0.60) &  1.050 (+0.010)(-0.010) & 0.993 (+0.010)(-0.010)  & 1       \\
        1761984258439478784 & HIP103025      &  6.30  (+0.82)(-0.29)  &  1.020 (+0.009)(-0.008)  &  1.130 (+0.009)(-0.009)  &   7.10 (+0.30)(-0.40) &  1.000 (+0.008)(-0.008) & 1.123 (+0.010)(-0.010)  & 1       \\
        990844114760515840  & HD49425        &  5.40  (+1.07)(-0.67)  &  0.990 (+0.009)(-0.008)  &  1.010 (+0.007)(-0.009)  &   5.60 (+0.40)(-0.60) &  0.973 (+0.015)(-0.010) & 0.998 (+0.008)(-0.008)  & 1       \\
        272026980672350336  & HIP20722       &  1.00  (+0.62)(-0.27)  &  1.070 (+0.017)(-0.009)  &  0.980 (+0.016)(-0.008)  &   0.90 (+0.50)(-0.40) &  1.065 (+0.010)(-0.010) & 0.973 (+0.010)(-0.008)  & 1       \\
        740735868326858368  & HIP49580       &  2.80  (+1.14)(-0.79)  &  1.010 (+0.014)(-0.007)  &  0.970 (+0.016)(-0.015)  &   2.60 (+0.70)(-0.50) &  1.008 (+0.010)(-0.013) & 0.955 (+0.008)(-0.010)  & 1       \\
        1508557582834745088 & HIP69709       &  6.90  (+0.91)(-0.67)  &  1.030 (+0.010)(-0.009)  &  1.180 (+0.016)(-0.023)  &   6.80 (+0.50)(-0.40) &  1.018 (+0.008)(-0.010) & 1.160 (+0.010)(-0.010)  & 1       \\
        3557158782894062080 & HIP52649       &  6.90  (+1.08)(-0.25)  &  1.000 (+0.008)(-0.009)  &  1.120 (+0.029)(-0.008)  &   7.80 (+0.40)(-0.30) &  0.978 (+0.008)(-0.008) & 1.115 (+0.010)(-0.010)  & 1       \\
        1260316891261098752 & HIP69230       &  1.19  (+1.29)(-0.57)  &  1.000 (+0.017)(-0.012)  &  0.900 (+0.008)(-0.007)  &   2.40 (+0.80)(-0.60) &  0.958 (+0.013)(-0.010) & 0.893 (+0.010)(-0.010)  & 1       \\
        3915429733361853952 & HIP55820       &  5.50  (+0.86)(-0.44)  &  0.960 (+0.016)(-0.009)  &  0.950 (+0.011)(-0.007)  &   5.70 (+0.40)(-0.30) &  0.940 (+0.008)(-0.010) & 0.940 (+0.008)(-0.008)  & 1       \\
        2136741591197747712 & HD184424       &  0.80  (+1.18)(-0.48)  &  1.060 (+0.008)(-0.008)  &  0.970 (+0.009)(-0.008)  &   0.80 (+0.60)(-0.60) &  1.060 (+0.010)(-0.010) & 0.965 (+0.010)(-0.010)  & 1       \\
        345765551891443968  & HIP9172        &  1.00  (+0.75)(-0.19)  &  1.030 (+0.017)(-0.010)  &  0.930 (+0.011)(-0.007)  &   0.70 (+0.40)(-0.40) &  1.025 (+0.010)(-0.010) & 0.920 (+0.010)(-0.008)  & 1       \\
        637329067477530368  & HIP46076       &  2.10  (+1.08)(-0.63)  &  1.000 (+0.008)(-0.008)  &  0.920 (+0.020)(-0.007)  &   2.90 (+0.70)(-0.90) &  0.968 (+0.018)(-0.013) & 0.913 (+0.010)(-0.008)  & 1       \\
        3203761005400205824 & HIP20218       &  2.50  (+0.91)(-0.44)  &  1.000 (+0.009)(-0.007)  &  0.940 (+0.008)(-0.008)  &   2.40 (+0.70)(-0.50) &  0.995 (+0.010)(-0.008) & 0.935 (+0.010)(-0.010)  & 1       \\
        1323563308352525440 & HIP79068       &  1.00  (+1.00)(-0.59)  &  1.080 (+0.014)(-0.008)  &  0.990 (+0.016)(-0.007)  &   0.60 (+0.40)(-0.30) &  1.078 (+0.008)(-0.010) & 0.980 (+0.008)(-0.008)  & 1       \\
        882410278028835584  & HIP38647       &  1.50  (+1.05)(-0.37)  &  1.010 (+0.011)(-0.008)  &  0.930 (+0.009)(-0.007)  &   1.30 (+0.70)(-0.60) &  1.008 (+0.010)(-0.010) & 0.920 (+0.010)(-0.008)  & 1       \\
        291701396922348672  & HIP7244        &  1.64  (+1.07)(-0.47)  &  1.000 (+0.013)(-0.007)  &  0.920 (+0.012)(-0.007)  &   1.80 (+0.60)(-0.60) &  0.988 (+0.010)(-0.010) & 0.925 (+0.008)(-0.010)  & 1       \\
        333106054183616000  & HIP10321       &  0.70  (+1.35)(-0.24)  &  1.010 (+0.014)(-0.009)  &  0.910 (+0.009)(-0.007)  &   1.60 (+0.60)(-0.60) &  0.995 (+0.010)(-0.010) & 0.910 (+0.008)(-0.010)  & 1       \\
        4511284599484268544 & HD229450A      &  6.30  (+0.82)(-0.36)  &  1.020 (+0.007)(-0.008)  &  1.100 (+0.008)(-0.008)  &   6.70 (+0.50)(-0.50) &  1.005 (+0.008)(-0.010) & 1.085 (+0.013)(-0.010)  & 1       \\
        2869747927140286208 & HD221103       &  2.80  (+0.70)(-0.27)  &  1.070 (+0.014)(-0.008)  &  1.050 (+0.008)(-0.008)  &   3.20 (+0.40)(-0.30) &  1.060 (+0.008)(-0.010) & 1.045 (+0.008)(-0.010)  & 1       \\
        2163433717746084352 & TYC3588-6756-1 &  0.60  (+1.21)(-0.29)  &  1.040 (+0.009)(-0.015)  &  0.940 (+0.010)(-0.007)  &   1.00 (+0.80)(-0.60) &  1.023 (+0.010)(-0.013) & 0.930 (+0.010)(-0.010)  & 1       \\
        4534307273451743744 & HD342979       &  1.30  (+1.49)(-0.70)  &  1.000 (+0.011)(-0.008)  &  0.910 (+0.009)(-0.007)  &   1.30 (+0.50)(-0.60) &  0.995 (+0.010)(-0.010) & 0.903 (+0.008)(-0.008)  & 1       \\
        1855968480863250176 & HIP101893      &  0.80  (+1.36)(-0.24)  &  1.030 (+0.014)(-0.019)  &  0.930 (+0.009)(-0.007)  &   1.20 (+0.60)(-0.60) &  1.013 (+0.010)(-0.010) & 0.920 (+0.010)(-0.008)  & 1       \\
        1909471786118933120 & HD213244       &  1.20  (+1.44)(-0.64)  &  1.000 (+0.008)(-0.009)  &  0.900 (+0.010)(-0.007)  &   1.00 (+0.60)(-0.40) &  0.988 (+0.008)(-0.010) & 0.890 (+0.008)(-0.010)  & 1       \\
        2196984318618671360 & TYC4250-1364-1 &  2.20  (+0.67)(-0.37)  &  1.030 (+0.016)(-0.007)  &  0.970 (+0.014)(-0.010)  &   1.50 (+0.50)(-0.60) &  1.030 (+0.010)(-0.008) & 0.948 (+0.010)(-0.010)  & 1       \\
        1792706331305783808 & TYC1678-109-1  &  2.30  (+1.50)(-0.86)  &  1.010 (+0.009)(-0.009)  &  0.950 (+0.016)(-0.009)  &   2.20 (+0.50)(-0.60) &  1.003 (+0.010)(-0.010) & 0.935 (+0.010)(-0.010)  & 1       \\
        1962597885872344704 & TYC3210-873-1  &  1.30  (+1.13)(-0.31)  &  1.030 (+0.017)(-0.010)  &  0.940 (+0.009)(-0.007)  &   1.00 (+0.50)(-0.50) &  1.033 (+0.010)(-0.010) & 0.935 (+0.010)(-0.008)  & 1       \\
        2003130042007448704 & HD235929       &  0.90  (+1.26)(-0.26)  &  0.990 (+0.016)(-0.012)  &  0.890 (+0.008)(-0.008)  &   0.70 (+0.80)(-0.50) &  0.975 (+0.010)(-0.010) & 0.873 (+0.010)(-0.008)  & 1       \\
        1796825582900925056 & TYC2211-2123-1 &  3.90  (+0.87)(-0.39)  &  1.030 (+0.014)(-0.009)  &  1.010 (+0.009)(-0.007)  &   3.70 (+0.50)(-0.60) &  1.025 (+0.008)(-0.010) & 1.003 (+0.010)(-0.010)  & 1       \\
        1957736635726467840 & TYC3208-1984-1 &  2.30  (+0.92)(-0.53)  &  1.010 (+0.009)(-0.009)  &  0.940 (+0.014)(-0.008)  &   1.60 (+0.40)(-0.60) &  1.010 (+0.010)(-0.010) & 0.923 (+0.008)(-0.010)  & 1       \\
        1989930095677359616 & HD235957       &  8.10  (+0.53)(-0.25)  &  0.970 (+0.014)(-0.009)  &  1.070 (+0.014)(-0.007)  &   8.00 (+0.50)(-0.40) &  0.958 (+0.008)(-0.010) & 1.053 (+0.010)(-0.008)  & 1       \\
        225038148667264384  & HD22875        &  1.70  (+1.04)(-0.67)  &  1.010 (+0.008)(-0.008)  &  0.920 (+0.009)(-0.007)  &   1.60 (+0.70)(-0.60) &  0.995 (+0.010)(-0.010) & 0.913 (+0.010)(-0.008)  & 1       \\
        1807690750662760576 & HD190340       &  3.70  (+0.75)(-0.36)  &  1.070 (+0.009)(-0.009)  &  1.080 (+0.014)(-0.008)  &   4.10 (+0.40)(-0.50) &  1.053 (+0.010)(-0.010) & 1.070 (+0.010)(-0.010)  & 1\dag   \\
        2592050460064616832 & HD9201         &  3.60  (+1.06)(-0.65)  &  1.020 (+0.014)(-0.009)  &  0.990 (+0.008)(-0.009)  &   3.10 (+0.60)(-0.40) &  1.015 (+0.010)(-0.010) & 0.978 (+0.008)(-0.010)  & 1       \\
        132371961511017856  & HIP11253       &  6.90  (+0.81)(-0.40)  &  1.010 (+0.015)(-0.008)  &  1.130 (+0.018)(-0.014)  &   7.50 (+0.50)(-0.40) &  1.000 (+0.010)(-0.008) & 1.153 (+0.020)(-0.018)  & 1\dag   \\
        1605622057017876992 & HIP71989       &  0.37  (+0.47)(-0.23)  &  1.050 (+0.014)(-0.007)  &  0.940 (+0.015)(-0.007)  &   0.30 (+0.40)(-0.30) &  1.048 (+0.008)(-0.010) & 0.943 (+0.008)(-0.008)  & 1\dag   \\
        2257140386478701568 & HIP91210       &  1.30  (+0.96)(-0.42)  &  1.060 (+0.015)(-0.009)  &  0.980 (+0.008)(-0.007)  &   1.40 (+0.50)(-0.50) &  1.053 (+0.010)(-0.008) & 0.975 (+0.010)(-0.008)  & 1\dag   \\
        3827020717791837056 & HIP47312       &  3.90  (+1.01)(-0.34)  &  1.040 (+0.009)(-0.010)  &  1.050 (+0.014)(-0.007)  &   4.70 (+0.40)(-0.50) &  1.023 (+0.008)(-0.010) & 1.053 (+0.010)(-0.010)  & 1\dag   \\
        855523714036230016  & HIP52278       &  2.10  (+0.99)(-0.40)  &  0.990 (+0.007)(-0.008)  &  0.910 (+0.010)(-0.007)  &   3.40 (+0.50)(-0.50) &  0.950 (+0.010)(-0.008) & 0.903 (+0.010)(-0.010)  & 1\dag   \\
        3574377616021489024 & HD105590A      &  4.50  (+1.07)(-0.45)  &  1.000 (+0.009)(-0.009)  &  1.000 (+0.015)(-0.009)  &   5.40 (+0.30)(-0.60) &  0.980 (+0.010)(-0.008) & 0.995 (+0.010)(-0.008)  & 1\dag   \\
        1610658885425227264 & HIP71440       &  7.40  (+1.09)(-0.53)  &  1.000 (+0.010)(-0.009)  &  1.150 (+0.020)(-0.031)  &   8.20 (+0.70)(-0.60) &  0.978 (+0.010)(-0.013) & 1.138 (+0.033)(-0.028)  & 1\dag   \\
        \hline
        \hline
    \end{tabular}
    \begin{tablenotes}
	\item \textbf{Notes.}(1) Solar twins, (2) Solar proxies, (3) Solar analogs, (4) Solar-type stars, (\dag) Binary star.
	\end{tablenotes}
\end{table}
\end{landscape}

\begin{table}
    \centering
    \caption{Solar twin stars used to calibrated our $S_{\rm{HK, TS23}}$ indices into the Mount Wilson system (MW).}
    \label{appendix_tab:3}
    \begin{tabular}{cccccc}
        \hline
        \hline
          ID      &  $S_{\rm{HK, TS23}}$ & $\sigma(S_{\rm{HK, TS23}})$ &  $S_{\rm{MW}}$ & $\sigma(S_{\rm{MW}})$  &  Reference  \\
        \hline                                                                                                    
        HIP7585   &          0.128       &           0.003             &     0.177      &       0.004            &  ($\star$)  \\
        HIP49756  &          0.118       &           0.002             &     0.163      &       0.002            &  ($\star$)  \\
        HIP79672  &          0.125       &           0.001             &     0.167      &       0.004            &  ($\star$)  \\
        HIP95962  &          0.113       &           0.001             &     0.163      &       0.002            &  ($\star$)  \\
        HIP8507   &          0.129       &           0.005             &     0.174      &       0.005            &  ($\star$)  \\
        HIP77052  &          0.143       &           0.000             &     0.214      &       0.015            &  ($\star$)  \\
        HIP28066  &          0.117       &           0.001             &     0.158      &       0.001            &  ($\star$)  \\
        HIP85042  &          0.112       &           0.001             &     0.158      &       0.004            &  ($\star$)  \\
        HIP118115 &          0.111       &           0.001             &     0.157      &       0.001            &  ($\star$)  \\
        HIP102040 &          0.129       &           0.000             &     0.176      &       0.005            &  ($\star$)  \\
        HIP113357 &          0.109       &           0.004             &     0.154      &       0.001            &  (\dag)     \\
        HIP8102   &          0.127       &           0.001             &     0.171      &       0.001            &  (\dag)     \\
        HIP1499   &          0.113       &           0.002             &     0.157      &       0.002            &  (\dag)     \\
        HIP59532  &          0.109       &           0.005             &     0.158      &       0.002            &  (\dag)     \\
        HIP106006 &          0.114       &           0.001             &     0.158      &       0.003            &  (\dag)     \\
        \hline
        \hline
    \end{tabular}
    \begin{tablenotes}
	\item \textbf{Notes}. ($\star$) \citet{Diego:2018A&A...619A..73L}; (\dag) this work.
	\end{tablenotes}
\end{table}

\begin{table}
    \centering
    \caption{Spectroscopic binary stars.}
    \label{appendix_tab:4}
    \begin{tabular}{cc}
        \hline
        \hline
        \textit{Gaia} EDR3   & Identifier       \\
        \hline              
        3985360665753530112 & HIP 54191        \\
        1438813773578253312 & TYC 4202-561-1   \\
        2642456024453117056 & HD 224033        \\
        302429645407269120  & TYC 1762-1034-1  \\
        \hline
        \hline
    \end{tabular}
\end{table}

\bsp	
\label{lastpage}
\end{document}